\documentclass[useAMS,usenatbib]{mn2e}
\usepackage{graphicx}

\def\msun{M_\odot}
\def\fs{f_{\rm S}}
\def\fsi{f_{\rm S_I}}
\def\fsv{f_{\rm S_V}}
\def\I0{I_{\rm 0}}
\def\IS{I_{\rm S}}
\def\VIS{(V-I)_{\rm S}}
\def\tE{t_{\rm E}}
\def\te{t_{\rm E}}
\def\t0{t_{\rm 0}}
\def\u0{u_{\rm 0}}

\newcommand{\eg}{{e.g.},\,}
\newcommand{\ie}{{i.e.},\,}
\newcommand{\apj}{{Astrophysical Journal}}
\newcommand{\apjl}{{Astrophysical Journal Letters}}
\newcommand{\apjs}{{Astrophysical Journal Supplement Series}}
\newcommand{\araa}{{ARA\&A}}
\newcommand{\aap}{{Astronomy \& Astrophysics}}
\newcommand{\mnras}{{MNRAS}}

\title[The OGLE-III View of Microlensing towards the LMC.]
{The OGLE View of Microlensing towards the Magellanic Clouds. III. Ruling out sub-solar MACHOs with the OGLE-III LMC data.\thanks{Based on
    observations obtained with the 1.3~m Warsaw telescope at the Las Campanas Observatory of the Carnegie Institution of Washington.}}
\author[{\L}. Wyrzykowski et al.]
{{\L}. Wyrzykowski$^{1,3}$\thanks{email: wyrzykow@ast.cam.ac.uk, name
    pronunciation: {\it Woocash Vizhikovsky}}, S. Koz{\l}owski$^{2,3}$,
  J. Skowron$^{2,3}$,  A. Udalski$^3$, M. K. Szyma{\'n}ski$^3$, \newauthor
M. Kubiak$^3$,  G. Pietrzy{\'n}ski$^{3,4}$, I. Soszy{\'n}ski$^3$, O. Szewczyk$^{3,4}$, \newauthor
K. Ulaczyk$^3$, R. Poleski$^3$\\
  $^1$ Institute of Astronomy, University of Cambridge,  Madingley~Road,  Cambridge~CB3~0HA,~UK \\
  $^2$ Department of Astronomy, The Ohio State University, 140 W. 18th Ave., Columbus, OH 43210, USA\\
  $^3$ Warsaw University Astronomical Observatory, Al.~Ujazdowskie~4, 00-478~Warszawa, Poland \\
  $^4$ Universidad de Concepci{\'o}n, Departamento de Fisica, Astronomy Group, Casilla 160-C, Concepci{\'o}n, Chile\\ }

\begin{document}

\date{Accepted 2010 December 3. Received 2010 December 3; in original form 2010 September 9.}

\pagerange{\pageref{firstpage}--\pageref{lastpage}} \pubyear{2010}

\maketitle

\label{firstpage}

\begin{abstract}

In the third part of the series presenting the Optical Gravitational Lensing Experiment (OGLE) microlensing studies of the dark matter halo compact objects (MACHOs) we describe results of the OGLE-III monitoring of the Large Magellanic Cloud (LMC).
This unprecedented data set contains almost continuous photometric coverage over 8 years of about 35 million objects spread over 40 square degrees.
We report a detection of two candidate microlensing events found with the automated pipeline and an additional two, less probable, candidate events found manually. 
The optical depth derived for the two main candidates was calculated following a detailed blending examination and detection efficiency determination and was found to be
$\tau=(0.16 \pm 0.12) \times 10^{-7}$. 

If the microlensing signal we observe originates from MACHOs it means their masses are around $0.2~\msun$ and they compose only $f=3\pm2$ per cent of the mass of the Galactic Halo. However, the more likely explanation of our detections does not involve dark matter compact objects at all and rely on natural effect of self-lensing of LMC stars by LMC lenses. 
In such a scenario we can almost completely rule out MACHOs in the sub-solar mass range with an upper limit at $f<7$ per cent reaching its minimum of $f<4$ per cent at $M=0.1~\msun$.
For masses around $M=10\msun$ the constraints on the MACHOs are more lenient with $f\sim20$ per cent. 
Owing to limitations of the survey there is no reasonable limit found for heavier masses, leaving only a tiny window of mass spectrum still available for dark matter compact objects.


\end{abstract}


\section{Introduction}
The Milky Way's halo is probably one of the least known parts of our Galaxy. 
Numerous recent detections of tidal debris leftover after close encounters between smaller dwarf galaxies and our giant spiral confirm the presence of cold dark matter (CDM) substructures in the halo (\eg \citealt{Belokurov2006Streams}). However, the following question still remains unanswered: what actually is the dark matter (DM) or is not.
Compact dark matter objects (MACHOs) would have been the most convenient explanation and in the last two decades this theory has been tested using various methods sensitive to different ranges of masses.
In the high mass regime ($M>30~\msun$) wide halo binary objects were studied, but no signature of disturbance due to MACHOs was detected (see \citealt{Yoo2004MACHO} and \citealt{Quinn2009}).

For detecting stellar-mass compact DM objects the technique of gravitational microlensing was suggested by \citet{Paczynski1986}.
It employs a unique feature of gravitational lensing, namely its sensitivity to unseen objects when they bend and amplify the light of a distant source. 
The idea was simple: let us observe some distant background rich in stars (\eg Magellanic Clouds) and wait for their temporal brightening due to passage of a massive object located along the line-of-sight between us and the source. 
Several observing campaigns started after Paczynski's proposal: MACHO \citep{MACHO}, OGLE \citep{Udalski1993}, EROS \citep{EROS} and MOA \citep{MOA}, which for many years observed the Large and Small Magellanic Clouds (LMC and SMC, respectively).

The MACHO collaboration was first to publish their results and claimed 20 per cent of mass of the galactic halo is composed of MACHOs 
with an average mass of 0.4 $\msun$
(\citealt{AlcockMACHOLMC}, \citealt{BennettMACHOLMC}). Based on detection of 10 candidates for microlensing events found in the central 15 sq. deg of the LMC observed over 5.7 years they derived the optical depth towards the LMC of $\tau_{\rm LMC} = (1.0 \pm 0.3) \times 10^{-7}$ \citep{BennettMACHOLMC}. 

In 2007 the EROS group analysed their data and published their conclusions, which contradicted the MACHO result quite severely. In their data comprising of 6.7 years of continuous observations of 84 deg$^2$ they found no candidates for microlensing events among their bright sample of stars \citep{TisserandEROSLMC}. 
This led them to the upper limit for the optical depth of $\tau_{\rm LMC} < 0.36 \times 10^{-7}$, which translated to $f = M_{\rm Macho}$/$M_{\rm halo} < 8$ per cent for MACHOs with masses $0.4~\msun$.
  
The OGLE project monitored the LMC during its second (1996-2000) and third (2001-2009) phases (hereafter OGLE--II and OGLE--III, respectively), and the observations are still carried on in the current OGLE--IV phase.
Our study of the OGLE--II data (\citealt{Wyrzykowski2009}, herafter Paper I) led to the detection of two candidates for microlensing events, OGLE-LMC-01 and OGLE-LMC-02, and $\tau_{\rm LMC} = 0.43\pm0.33\times 10^{-7}$, which is closer to the limit of EROS than the value of MACHO.
Moreover, all detected microlensing signal can be attributed to self-lensing (i.e., when foreground LMC stars microlens backgound LMC stars), leaving no room for lensing due to DM halo objects in the sub-solar and solar mass range.

However, the OGLE--II phase lasted only 4 years and covered only parts of the central bar of the LMC (about 4.7 deg$^2$), therefore its result is naturally limited and is subject to some uncertainty. Due to small number statistics, the upper limit on the MACHO presence in the Milky Way's halo was estimated only at $f<20$ per cent.

In this paper we present yet another voice in this turbulent story of microlensing towards the Magellanic Clouds and report the results of the search for microlensing events in the OGLE--III data gathered towards the LMC. OGLE--III overwhelms most previous studies in terms of its duration (8 years) and coverage (40 deg$^2$) 
but more importantly, it uniformly covers the entire LMC bar region and much of LMC outskirts. 

The paper is organised as follows. In Section \ref{sec:data}, the OGLE--III LMC dataset is described.
In the following section, the search algorithm is presented. Section \ref{sec:results} contains the results of the search and detailed description of all detected candidates for microlensing events.
Next, the blending study is described and the optical depth is derived. The paper finishes with discussion of the results and conclusions.

\begin{table*}
\centering
\caption{OGLE--III LMC fields.}
\label{tab:fields}
\begin{tiny}
\begin{tabular}{cccrrc|cccrrc}
\hline
\noalign{\vskip5pt}
Field & $\alpha_{J2000}$ & $\delta_{J2000}$ & \multicolumn{2}{c}{$N_{*\mathrm{good}}$[$10^3$]} & 
$\langle \log{N_{\mathrm{all_{CCD}}}}\rangle$ & 
Field & $\alpha_{J2000}$ & $\delta_{J2000}$ & \multicolumn{2}{c}{$N_{*\mathrm{good}}$[$10^3$]} & 
$\langle \log{N_{\mathrm{all_{CCD}}}}\rangle$ \\
& & & tpl & real &  & & & & tpl & real &  \\

\noalign{\vskip5pt}
\hline
\noalign{\vskip5pt}
LMC100 & 5:19:02.2 & -69:15:07 & 712.2 & 973.9 & 5.13 & LMC158 & 4:30:59.9 & -70:26:01 & 20.0 & 20.8 & 3.63 \\
LMC101 & 5:19:03.1 & -68:39:19 & 279.4 & 323.2 & 4.84 & LMC159 & 5:25:11.4 & -68:03:58 & 129.1 & 137.6 & 4.47 \\
LMC102 & 5:19:03.4 & -68:03:48 & 125.9 & 134.7 & 4.52 & LMC160 & 5:25:20.9 & -68:39:24 & 264.4 & 296.0 & 4.74 \\
LMC103 & 5:19:02.9 & -69:50:26 & 540.6 & 687.4 & 5.02 & LMC161 & 5:25:32.5 & -69:14:59 & 453.6 & 561.8 & 4.99 \\
LMC104 & 5:19:02.4 & -70:26:03 & 272.1 & 313.2 & 4.84 & LMC162 & 5:25:43.3 & -69:50:24 & 817.7 & 1159.9 & 5.17 \\
LMC105 & 5:19:01.6 & -71:01:31 & 205.0 & 224.2 & 4.65 & LMC163 & 5:25:52.2 & -70:25:55 & 524.6 & 669.5 & 5.00 \\
LMC106 & 5:19:01.0 & -71:36:57 & 135.5 & 144.6 & 4.47 & LMC164 & 5:26:08.4 & -71:01:23 & 197.5 & 217.1 & 4.68 \\
LMC107 & 5:13:01.5 & -66:52:57 & 93.2 & 99.3 & 4.45 & LMC165 & 5:26:20.9 & -71:37:01 & 140.6 & 152.3 & 4.61 \\
LMC108 & 5:13:01.9 & -67:28:40 & 105.8 & 113.2 & 4.51 & LMC166 & 5:31:20.1 & -68:03:51 & 127.8 & 138.5 & 4.59 \\
LMC109 & 5:12:53.3 & -68:04:06 & 138.8 & 149.4 & 4.57 & LMC167 & 5:31:39.6 & -68:39:32 & 180.9 & 200.1 & 4.69 \\
LMC110 & 5:12:43.6 & -68:39:42 & 293.4 & 335.6 & 4.82 & LMC168 & 5:32:01.4 & -69:15:00 & 316.2 & 374.0 & 4.90 \\
LMC111 & 5:12:32.7 & -69:15:02 & 441.2 & 526.7 & 4.94 & LMC169 & 5:32:22.8 & -69:50:26 & 757.6 & 1054.8 & 5.14 \\
LMC112 & 5:12:21.5 & -69:50:21 & 481.7 & 594.0 & 4.97 & LMC170 & 5:32:48.1 & -70:25:53 & 588.6 & 769.3 & 5.04 \\
LMC113 & 5:12:10.9 & -70:25:48 & 289.7 & 334.2 & 4.85 & LMC171 & 5:33:10.6 & -71:01:30 & 235.8 & 268.3 & 4.81 \\
LMC114 & 5:11:58.9 & -71:01:22 & 109.4 & 116.2 & 4.44 & LMC172 & 5:33:34.4 & -71:36:54 & 185.7 & 205.9 & 4.72 \\
LMC115 & 5:07:09.7 & -66:52:59 & 133.8 & 143.0 & 4.47 & LMC173 & 5:37:29.3 & -68:03:50 & 126.9 & 135.2 & 4.44 \\
LMC116 & 5:07:00.9 & -67:28:29 & 117.3 & 124.4 & 4.41 & LMC174 & 5:37:59.8 & -68:39:26 & 155.5 & 169.2 & 4.63 \\
LMC117 & 5:06:55.3 & -68:03:58 & 260.4 & 298.2 & 4.82 & LMC175 & 5:38:32.3 & -69:15:01 & 268.7 & 305.3 & 4.79 \\
LMC118 & 5:06:25.4 & -68:39:25 & 383.6 & 463.1 & 4.94 & LMC176 & 5:39:01.6 & -69:50:30 & 357.0 & 414.9 & 4.86 \\
LMC119 & 5:06:02.5 & -69:15:02 & 578.7 & 723.4 & 5.01 & LMC177 & 5:39:38.0 & -70:25:49 & 459.3 & 577.8 & 5.01 \\
LMC120 & 5:05:39.8 & -69:50:28 & 339.4 & 399.4 & 4.89 & LMC178 & 5:40:14.1 & -71:01:27 & 260.6 & 292.4 & 4.77 \\
LMC121 & 5:05:14.4 & -70:25:59 & 200.6 & 223.5 & 4.74 & LMC179 & 5:40:52.3 & -71:36:58 & 167.3 & 180.7 & 4.57 \\
LMC122 & 5:04:52.9 & -71:01:25 & 139.3 & 150.6 & 4.56 & LMC180 & 5:40:51.5 & -72:12:28 & 111.6 & 120.0 & 4.51 \\
LMC123 & 5:01:18.0 & -66:53:00 & 113.5 & 121.3 & 4.48 & LMC181 & 5:43:35.7 & -68:03:58 & 97.2 & 103.1 & 4.40 \\
LMC124 & 5:01:00.3 & -67:28:27 & 136.9 & 147.6 & 4.56 & LMC182 & 5:44:16.0 & -68:39:32 & 141.0 & 152.9 & 4.59 \\
LMC125 & 5:00:36.1 & -68:03:54 & 162.5 & 177.3 & 4.65 & LMC183 & 5:45:02.8 & -69:14:59 & 173.3 & 189.9 & 4.66 \\
LMC126 & 5:00:02.4 & -68:39:31 & 270.9 & 309.7 & 4.82 & LMC184 & 5:45:43.2 & -69:50:33 & 243.2 & 277.0 & 4.78 \\
LMC127 & 4:59:33.6 & -69:14:54 & 296.1 & 340.8 & 4.84 & LMC185 & 5:46:30.8 & -70:25:51 & 350.9 & 413.7 & 4.90 \\
LMC128 & 4:59:03.6 & -69:50:24 & 184.2 & 203.4 & 4.71 & LMC186 & 5:47:21.2 & -71:01:24 & 205.6 & 225.9 & 4.67 \\
LMC129 & 4:58:24.6 & -70:26:07 & 151.9 & 164.6 & 4.58 & LMC187 & 5:48:12.6 & -71:36:52 & 141.6 & 154.2 & 4.57 \\
LMC130 & 4:57:50.8 & -71:01:20 & 118.1 & 126.2 & 4.46 & LMC188 & 5:48:26.6 & -72:12:27 & 68.0 & 71.9 & 4.28 \\
LMC131 & 4:55:28.6 & -66:52:46 & 133.3 & 142.8 & 4.49 & LMC189 & 5:50:37.9 & -68:39:26 & 81.7 & 86.1 & 4.28 \\
LMC132 & 4:55:00.6 & -67:28:36 & 107.1 & 114.0 & 4.44 & LMC190 & 5:51:33.2 & -69:14:55 & 107.7 & 114.4 & 4.40 \\
LMC133 & 4:54:29.2 & -68:03:47 & 186.6 & 204.2 & 4.64 & LMC191 & 5:52:20.1 & -69:50:24 & 137.7 & 147.9 & 4.48 \\
LMC134 & 4:53:49.2 & -68:39:18 & 158.5 & 171.4 & 4.58 & LMC192 & 5:53:24.1 & -70:25:51 & 143.3 & 154.0 & 4.46 \\
LMC135 & 4:53:05.2 & -69:14:51 & 140.0 & 150.6 & 4.55 & LMC193 & 5:54:21.7 & -71:01:34 & 83.4 & 88.1 & 4.24 \\
LMC136 & 4:52:23.7 & -69:50:25 & 117.3 & 125.1 & 4.49 & LMC194 & 5:55:29.7 & -71:36:59 & 44.8 & 46.8 & 3.99 \\
LMC137 & 4:51:30.2 & -70:26:01 & 87.0 & 92.3 & 4.40 & LMC195 & 5:56:00.0 & -72:12:25 & 24.9 & 25.8 & 3.73 \\
LMC138 & 4:49:34.7 & -66:53:07 & 52.2 & 54.8 & 4.17 & LMC196 & 5:56:54.7 & -68:39:29 & 51.7 & 54.2 & 4.16 \\
LMC139 & 4:49:05.2 & -67:28:30 & 52.8 & 55.5 & 4.21 & LMC197 & 5:58:02.7 & -69:15:06 & 50.2 & 52.5 & 4.06 \\
LMC140 & 4:48:18.2 & -68:04:05 & 81.9 & 86.8 & 4.37 & LMC198 & 5:59:02.5 & -69:50:35 & 43.0 & 44.8 & 3.95 \\
LMC141 & 4:47:26.7 & -68:39:36 & 85.6 & 90.7 & 4.38 & LMC199 & 6:00:14.7 & -70:26:00 & 39.2 & 40.8 & 3.90 \\
LMC142 & 4:46:31.9 & -69:15:08 & 112.7 & 120.4 & 4.45 & LMC200 & 6:01:27.5 & -71:01:36 & 35.1 & 36.6 & 3.87 \\
LMC143 & 4:45:43.1 & -69:50:19 & 78.7 & 82.9 & 4.29 & LMC201 & 6:02:45.9 & -71:37:04 & 46.9 & 49.2 & 4.06 \\
LMC144 & 4:44:40.2 & -70:26:01 & 55.8 & 58.5 & 4.16 & LMC202 & 6:03:28.3 & -72:12:34 & 42.9 & 44.9 & 4.00 \\
LMC145 & 4:43:47.5 & -66:52:43 & 30.1 & 31.3 & 3.91 & LMC203 & 6:03:29.9 & -72:48:04 & 40.9 & 42.8 & 3.95 \\
LMC146 & 4:43:03.0 & -67:28:17 & 39.4 & 41.2 & 4.03 & LMC204 & 6:03:14.6 & -68:39:25 & 55.3 & 58.1 & 4.13 \\
LMC147 & 4:42:07.8 & -68:03:55 & 46.9 & 49.1 & 4.11 & LMC205 & 6:04:32.9 & -69:15:04 & 36.7 & 38.4 & 4.01 \\
LMC148 & 4:41:06.8 & -68:39:27 & 49.1 & 51.4 & 4.14 & LMC206 & 6:05:40.3 & -69:50:27 & 38.3 & 40.0 & 3.99 \\
LMC149 & 4:40:05.1 & -69:14:57 & 52.3 & 54.8 & 4.16 & LMC207 & 6:07:04.2 & -70:25:55 & 35.9 & 37.4 & 3.95 \\
LMC150 & 4:39:05.3 & -69:50:16 & 44.8 & 46.8 & 4.08 & LMC208 & 6:08:30.4 & -71:01:27 & 44.1 & 46.2 & 4.04 \\
LMC151 & 4:37:51.6 & -70:25:45 & 38.1 & 39.8 & 4.05 & LMC209 & 6:10:07.0 & -71:37:00 & 37.5 & 39.1 & 3.91 \\
LMC152 & 4:37:54.1 & -66:52:52 & 25.3 & 26.5 & 3.76 & LMC210 & 6:10:55.7 & -72:12:37 & 34.5 & 36.1 & 3.94 \\
LMC153 & 4:37:01.7 & -67:28:30 & 27.3 & 28.5 & 3.85 & LMC211 & 6:11:22.0 & -72:48:04 & 31.3 & 32.7 & 3.88 \\
LMC154 & 4:35:59.1 & -68:04:02 & 26.6 & 27.8 & 3.91 & LMC212 & 6:11:04.0 & -69:14:50 & 39.9 & 41.7 & 4.01 \\
LMC155 & 4:34:49.4 & -68:39:32 & 33.0 & 34.5 & 3.95 & LMC213 & 6:12:17.9 & -69:50:37 & 32.7 & 34.1 & 3.81 \\
LMC156 & 4:33:32.7 & -69:15:00 & 34.3 & 35.9 & 3.96 & LMC214 & 6:13:58.2 & -70:26:08 & 31.6 & 32.9 & 3.88 \\
LMC157 & 4:32:23.8 & -69:50:26 & 25.6 & 26.6 & 3.74 & LMC215 & 6:15:36.4 & -71:01:28 & 32.0 & 33.3 & 3.88 \\
\hline
\noalign{\vskip5pt}
\multicolumn{9}{r}{total}      & 19,424.4 & 22,740.0 & \\
\noalign{\vskip5pt}
\hline
\end{tabular}
\end{tiny}
\medskip
\begin{flushleft}
{\it Note:} Coordinates point to the centre of the field (centre of the mosaic), each being $35' \times
35'$. Number of ``good'' objects in the template is provided ($N>80$ and
$\langle I \rangle < 20.4$ mag) together with the estimated number of
real monitored stars (see Section \ref{sec:blending}). 
Mean number of all objects detected on a single CCD used for calculating the density of a field is given in the last column.
\end{flushleft}
\medskip
\bigskip
\bigskip
\end{table*}

\begin{figure*}
\includegraphics[width=12cm]{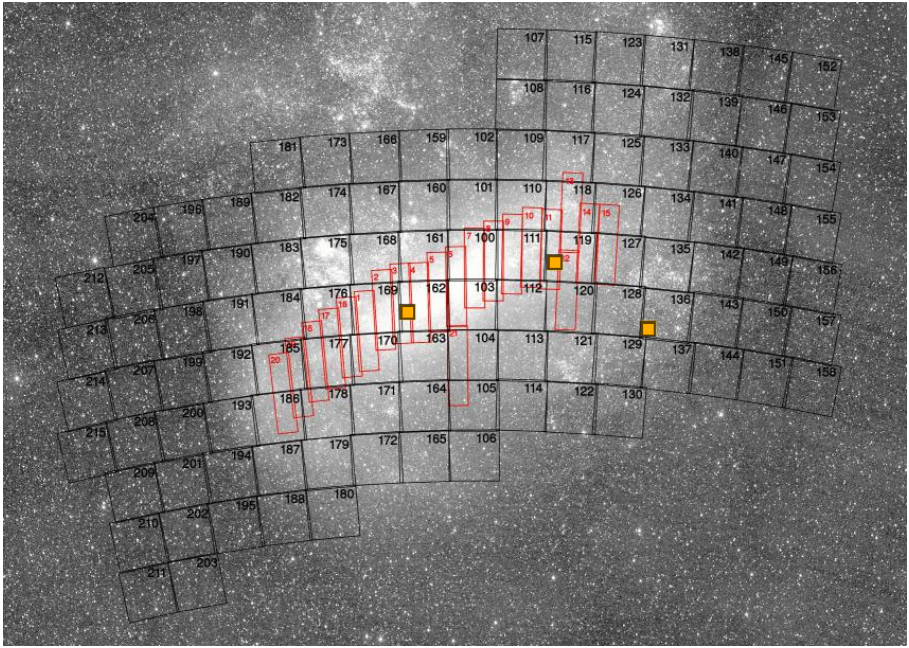}
\caption{Positions of the OGLE--III LMC fields (black). Also shown are all OGLE--II fields (red rectangles).
  The three small filled squares show  the positions of the HST fields used for our blending determination.
  Background image credit: ASAS all sky survey.}
\label{fig:fields}
 \end{figure*}

\section{Observational Data}
\label{sec:data}

The photometric observations presented in this work were carried out during the third phase of the OGLE project (2001-2009) 
with the 1.3-m Warsaw telescope located at Las Campanas Observatory,
Chile, operated by the Carnegie Institution of Washington. 
The ``second generation'' camera comprised of eight SITe
$2048\times4096$ CCD detectors with 15~$\mu$m pixels resulting in
0.26~arcsec/pixel scale and $35\times35$~arcmins field of view. 
The details on the instrumentation setup can be found in \citep{Udalski2003}.

There were 116 fields in the LMC covering a total of 40 square degrees. 
Their distribution on the sky is shown in Fig. \ref{fig:fields}.
The fields are listed in Table \ref{tab:fields} along with the coordinates of their centers, 
the number of ``good'' template objects in the $I$-band, the blending-corrected number of stars
(see Section \ref{sec:blending}) and the mean number of all objects visible on a single CCD (of 8) used for deriving the blending density level.
By ``good'' we mean those template objects which have at least 80 good data points 
(excluding measurements with very large error-bars) and mean magnitude brighter than $20.4$ mag.
The limiting magnitude was selected based on the mean observed luminosity function.

The very first observations of the LMC within the OGLE--III phase were taken in July 2001 (JD=$2\;452\;115.9$), however, regular monitoring started in September 2001 and continued until May 2009 (JD=$2\;454\;964.5$).
The vast majority of observations were done through the Cousins $I$-band filter with exposure time of 180~s. 
Between 385 and 637 measurements were taken in each field with an average sampling varying from 3.0 to 4.6 days between subsequent frames (excluding the gaps between the seasons).
In addition to that, between 30 and 137 observations per field were obtained in Johnson $V$-band and integration time of 225~s.
Average sampling frequency in the $V$-band was between 4.2 and 13.6 days.

The images were reduced with the pipeline based on the Difference Image Analysis (DIA; \citealt{AlardLuptonDIA}, \citealt{WozniakDIA}).
The photometry used in this work comes from the final reduction calibrated to the standard system.
Full description of the reduction techniques, photometric calibration and astrometric transformations can be found in \citep{Udalski2008OGLE3}.
 
Photometric errors produced by the DIA package were corrected following the method similar to that described in Paper I.
For each field the constant stars were selected based on their low photometric dispersion. 
These stars were then used for deriving the magnitude dependent correction factor for the error-bars which minimized the difference between observed scatter of the data points and the error-bar.
For each field and filters $f=I,~V$ we derived parameters $\gamma_f$ and $\epsilon_f$, which correct the original error-bar with the formula:
\begin{equation}
\sigma_{\rm mag_f~cor} = \sqrt{ ( \gamma_f \sigma_{\rm mag_f})^2 + \epsilon_f^2 },
\label{eq:errors}
\end{equation}
where $\sigma_{\rm mag}$ is the original error-bar returned by the photometry pipeline.

For all the LMC fields the mean values of the error-correction parameters were: 
$\langle\gamma_I\rangle = 1.2039$, 
$\langle\epsilon_I\rangle = 0.0046$, 
$\langle\gamma_V\rangle= 0.9956$,
$\langle\epsilon_V\rangle = 0.0035$.

As a side product of the error correction study we obtained also a formula for calculating error-bars of synthetic $I$-band magnitudes used in the light curves' simulations:

\begin{scriptsize}
\begin{eqnarray}
\Delta I_{\rm sim} = \Delta \, I_{\rm ref} 10^{-0.345 ( I_{\rm sim} - I_{\rm ref} ) + 0.019 ( I_{\rm sim}^2 - I_{\rm ref}^2 ) }, 
& {\rm for}~I_{\rm sim} \ge 15.0 & \\
\nonumber
\Delta I_{\rm sim} = 0.003, & {\rm for}~I_{\rm sim} < 15.0, & 
\label{eq:simerr}
\end{eqnarray}
\end{scriptsize}
where $I_{\rm sim}$ is the simulated magnitude for which the error bar ($\Delta
I_{\rm sim}$) is required, $I_{\rm ref}$ and $\Delta I_{\rm ref}$ are the
magnitude and the error bar of the reference star at a given epoch.
Such calculated error bars still need to be corrected with eq. (\ref{eq:errors}).

Error corrections for $I$ and $V$ passbands for the first couple of fields are gathered in Table \ref{tab:errorcor}. The full table is available on-line on the OGLE website\footnote{http://ogle.astrouw.edu.pl/}.

\begin{table}
\centering
\caption{Error correction coefficients for each CCD chip of the first four OGLE--III LMC fields for $I$- and $V-$ bands. The full table is available on-line from the OGLE website.}
\label{tab:errorcor}
\begin{tiny}
\begin{tabular}{ccccc}
\hline
Field & $\gamma_\mathrm{I}$ & $\epsilon_\mathrm{I}$ & $\gamma_\mathrm{V}$ & $\epsilon_\mathrm{V}$ \\
\hline
LMC100.1 & 1.049373 & 0.00412404 & 0.917834 & 0.0035 \\
LMC100.2 & 1.060207 & 0.00485469 & 0.985456 & 0.0035 \\
LMC100.3 & 1.077131 & 0.00463489 & 1.006376 & 0.0035 \\
LMC100.4 & 1.135749 & 0.0039967 & 1.014130 & 0.0035 \\
LMC100.5 & 1.131603 & 0.00512842 & 0.976961 & 0.0035 \\
LMC100.6 & 1.119252 & 0.00442105 & 0.971518 & 0.0035 \\
LMC100.7 & 1.093156 & 0.00463935 & 0.956374 & 0.0035 \\
LMC100.8 & 1.075116 & 0.00473159 & 0.977142 & 0.0035 \\
LMC101.1 & 0.886630 & 0.00412372 & 0.913540 & 0.0035 \\
LMC101.2 & 0.924936 & 0.00485499 & 0.985448 & 0.0035 \\
LMC101.3 & 0.927672 & 0.00463465 & 0.990697 & 0.0035 \\
LMC101.4 & 0.961909 & 0.00399673 & 0.987186 & 0.0035 \\
LMC101.5 & 0.945618 & 0.00512809 & 0.900159 & 0.0035 \\
LMC101.6 & 0.958065 & 0.00442147 & 0.970733 & 0.0035 \\
LMC101.7 & 0.933999 & 0.00463917 & 0.975772 & 0.0035 \\
LMC101.8 & 0.925774 & 0.00473256 & 0.956004 & 0.0035 \\
LMC102.1 & 1.077252 & 0.00412372 & 1.018413 & 0.0035 \\
LMC102.2 & 1.067985 & 0.00485506 & 1.029716 & 0.0035 \\
LMC102.3 & 1.101820 & 0.00463536 & 1.018757 & 0.0035 \\
LMC102.4 & 1.166521 & 0.00399767 & 1.055142 & 0.0035 \\
LMC102.5 & 1.147459 & 0.00512799 & 0.989320 & 0.0035 \\
LMC102.6 & 1.104168 & 0.00442109 & 1.031484 & 0.0035 \\
LMC102.7 & 1.085202 & 0.00463924 & 0.978987 & 0.0035 \\
LMC102.8 & 1.107776 & 0.00473242 & 0.992293 & 0.0035 \\
LMC103.1 & 0.940445 & 0.00412385 & 0.916057 & 0.0035 \\
LMC103.2 & 0.941654 & 0.00485423 & 0.979160 & 0.0035 \\
LMC103.3 & 0.932683 & 0.0046345 & 0.953727 & 0.0035 \\
LMC103.4 & 0.949726 & 0.0039974 & 0.929897 & 0.0035 \\
LMC103.5 & 0.930998 & 0.00512794 & 0.905821 & 0.0035 \\
LMC103.6 & 0.945777 & 0.00442151 & 0.951769 & 0.0035 \\
LMC103.7 & 0.930089 & 0.00463928 & 0.976390 & 0.0035 \\
LMC103.8 & 0.922839 & 0.00473232 & 0.986609 & 0.0035 \\
\dots & & & &  \\
\hline
\end{tabular}
\end{tiny}
\end{table}

\section{Search procedure}
\label{sec:search}

The main search for microlensing event candidates was performed on the regular database containing light curves of all stars which were detected on the template images. 

The search criteria are gathered in Table \ref{tab:conditions} along with the number of objects left after each cut for two star samples: ``All Stars'' and ``Bright Stars''. 
We performed the search in two samples to allow for a comparison with the previous optical depth determinations by EROS (bright stars) and MACHO (all stars).
The samples are formed from ``good'' objects found on a template image and differ with magnitude cut (dubbed ``cut 0'') applied prior the actual search for events. 
The All Stars Sample consists of all objects down to 20.4 mag and the Bright Stars Sample has the cut 0 set to a Red Clump centre magnitude plus 1 mag, following \cite{TisserandEROSLMC} and Paper I.
The conditions follow the ones we derived for and presented in the study of OGLE--II LMC (Paper I) and SMC (\citealt{OGLE2SMC}, hereafter Paper II). 
Because of slightly different properties of the OGLE--III data and sampling, the search parameters were derived again through the Monte Carlo simulations. 
Random constant (\ie non-variable) stars from the database were artificially microlensed and run through the search pipeline to fine tune the parameters of the pipeline.

From the entire LMC OGLE--III database (about 35.5 million objects detected on a template) we selected 19.5 million and 5.8 million in All and Bright Stars Sample, respectively, which satisfied the condition to be a ``good'' object (cut 0a and 0b, respectively).
Light curves of these stars were then subjects to check for a positive consistent deviation over some baseline (cut 1). We flagged a star as having a bump when its summary significance (defined after \citealt{SumiOGLEBulge}) over a peak was larger than 30. It limited the sample to 5,529 objects (4,553 in Bright Sample) and included various types of outbursting variable stars like supernovae, novae, dwarf novae. Among these were also artefactual symmetric bumps with duration of couple of months caused by the moving light echo of the Supernova 1987a. All objects within 15 arminutes radius from the SN1987a remnant ($\alpha=05$:35:$28.01$, $\delta=-69$:16:$11.6$) were excluded in cut 2. This cut affected only parts of the Eastern CCD chips of field LMC168 and Western chips of LMC175.

Next, we removed all bright and blue objects residing in the so called ``Blue Bumper'' region of the colour-magnitude diagram (cut 3). These evasive variable stars are a well known problem in the searches for microlensing events (\eg \citealt{AlcockMACHOLMC}) as their light curves often exhibit brightening episodes similar to low-amplification microlensing events. However, most of them reveal their true nature with another bump occurring years after the first one, the feature which is extremely rare for genuine microlensing events (\eg \citealt{SkowronRepeating}). 
Note, the number of potential ``Blue Bumpers" removed from All and Bright Stars Samples differs. This is because in cut 3, before applying the colour-magnitude cut, all stars for which $V$-band photometry (hence the colour) was not available, were removed.

In the series of the following cuts we used parameters of the standard ``Paczy{\'n}ski'' microlensing model fit to the light curves. The model is described as:

\begin{equation}
\label{eq:I}
I = \I0 - 2.5\log\left[ \fs A + (1-\fs)\right],
\end{equation}

\noindent
where, $\I0$ is the baseline magnitude in the $I$ band and
$\fs$ is the blending fraction (ratio of lensed source flux to total blends' flux in the $I$ band). The microlensing amplification $A$ equals:

\begin{equation}
\label{eq:A}
A= { u^2 + 2 \over u\sqrt{u^2+4} } \qquad {\rm and} \qquad u= \sqrt{\u0^2 + {{(t-\t0)^2} \over {\tE^2}}}.
\end{equation}

\noindent
where, $\t0$ is the time of the maximum of the peak, $\tE$ is the Einstein radius crossing time (event's time-scale) and $\u0$ is
the event's impact parameter.
We fitted the light curves with blending parameter being fixed $\fs=1$ (no blending) and being free. Hereafter, non-blended (4 parameters) models are denoted with the subscript $\mu 4$. 

In the cut 4 we selected objects with a bump, described by 4-parameter microlensing fit, which was more significant than a constant line with its noise and scatter. It narrowed down our sample to only 488 (192) objects. Then, we requested there were at least 6 data points in the bump in range of 1 Einstein radius crossing time on both sides from the peak (cut 5).

Next, we compared the microlensing model fit with a supernova (SN) model fit, approximated by a composition of two exponents (cut 6). In our data set, covering 40 square degrees over 8 years, we should expect to detect about 30 SNe (following \eg \citealt{AlcockMACHOLMC}), assuming mean efficiency of SNe detection of 20 per cent. A visual inspection of light curves of the objects surviving cuts 0--5, showed numerous SNe, often with the background galaxy clearly visible on the finding chart. Most of them were located away from the main bar of the LMC in the most sparse fields, where the internal LMC extinction does not reduce their visibility and the stellar density is low. Light curves removed in the cut 6 were clearly asymmetrical and, apart from the obvious SNe, belonged to various kinds of outbursting variables, like novae or redder Be stars. 

In the remaining cuts we narrowed down our sample of candidates for microlensing events directly using derived parameters, like $\t0, \tE, \fs, \u0$ (cuts 7,8,10,11) and the goodness of fit (global and at the peak) (cut 9). As a result the search pipeline returned the same 4 candidates for microlensing events in both All Stars and Bright Stars Samples.

One of these objects, LMC164.3.892 ($\alpha_{J2000}=$5:26:33.88, $\delta_{J2000}=$-70:57:44.8), was cross-matched with previously known microlensing event candidate, EROS-LMC-1.
That event was reported to have a second episode of brightening after nearly 5 years \citep{TisserandEROSLMC} and in our data it exhibited the third one after another 7 years. It was therefore rejected as not a genuine microlensing event.

Another object, LMC100.6.50890 ($\alpha_{J2000}=$5:18:33.24, $\delta_{J2000}=$-69:11:09.5), had a long term feature, resembling half of a long time-scale event with amplitude of 0.5~mag, however, when combined with available OGLE--II data, it turned out to be an asymmetric Be star.

This leaves us with only two candidates for microlensing events. They were dubbed OGLE-LMC-03 and OGLE-LMC-04, respectively, continuing the numbering of OGLE LMC events started with the findings in the OGLE--II data \citep{Wyrzykowski2009}.

We also visually inspected a few hundred light curves with high signal to noise ratios, not limited by the magnitude cut, and discovered two additional potential candidates, named OGLE-LMC-05 and OGLE-LMC-06. 
These did not pass through our main search pipeline due to the mean magnitude below the 20.4 mag threshold (OGLE-LMC-05) or anomalous shape of the light curve (OGLE-LMC-06).

\begin{table*}
\centering
\caption{Selection criteria for search for microlensing events in the OGLE--III LMC data and number of objects left after each cut for the All Stars and Bright Stars Samples.}
\label{tab:conditions}
\begin{tabular}[h]{c|lrrr}
\hline
Cut no. & & & \multicolumn{2}{c}{No. of objects left} \\
            & & & All  &  Bright \\
\hline     
0a & Selection of ``good'' objects & $N>80$, $\langle I \rangle \le 20.4$ mag  &  19,424,384 & \\
0b &                                          & $N>80$, $\langle I \rangle \le 18.8$ mag &                  & 5,782,733 \\
& & & &\\

1 & Significant bump over baseline & $\displaystyle\sum_\mathrm{peak} \sigma_i > 30.0 $ & 5,529 & 4,553\\
& & & & \\

2 & SN1987a light echo filter & $| \overline{\xi} - \overline{\xi}_\mathrm{SN1987a} | < 15' $ & 5,413 & 4,466\\
& & & & \\

3 & ``Bumper'' cut$^\dagger$ & $\langle I \rangle>18.5$~mag,  $\langle V-I \rangle>0.5$~mag & 1,871 & 1,168 \\ 
& & & & \\

4 & Microlensing fit better than constant line fit & ${{\chi^2_\mathrm{line}-\chi^2_{{\mu} 4}}\over{{\chi^2_{{\mu}4}\over{N_{\mathrm{dof},\mu4}}}\sqrt{2N_{\mathrm{dof},\mu4,\mathrm{peak}}}}} > 140$  & 488 & 192\\
& & & & \\

5 & Number of points at the peak$^{\ast}$ & $N_\mathrm{peak} > 5$  & 478 & 184\\ 
& & & & \\

6 & Microlensing fit better than supernova fit, & $\chi^2_{SN} > MIN(\chi^2,\chi^2_{{\mu}4})$ & 302 & 126 \\
& & & & \\

7 & Peak within the data span   & $2115 \le {\t0} \le 4965$   & 284 & 114 \\
& [HJD-2450000]    &         &         \\
& & & & \\

8 & Blended fit converged & $0< \fs < 1.2$ & 88 & 41 \\
& & & & \\

9 & Conditions on goodness of microlensing fit &  ${{\chi^2}\over{N_\mathrm{dof}}} \le 2.6 $ and ${{\chi_{\mu4,\mathrm{peak}}^2}\over{N_{\mathrm{dof},\mu4,\mathrm{peak}}}} \le 4.5 $ & 7 & 6\\
& (global and at the peak) & & \\
& & & & \\

10 & Time-scale cut & $1 \le {\tE} \le 1000$ & 5 & 5 \\
& $[d]$ &  & \\
& & & & \\

11 & Impact parameter cut &  $0 < {\u0} \le 1 $ & 4(2)$^{\star}$ & 4(2)$^{\star}$ \\
& & & & \\

\hline
\end{tabular}
\\
\begin{flushleft}
$^{\dagger}$ magnitudes as in the field LMC100.1 (shifted according to the position of the center of Red Clump) \\
$^{\ast}$in the range of $\t0 \pm 1 \tE$ \\
$^{\star}$Two events were rejected based on additional data from EROS and OGLE--II, see text\\
\end{flushleft}
\end{table*}

\begin{figure*}
\includegraphics[width=10.0cm]{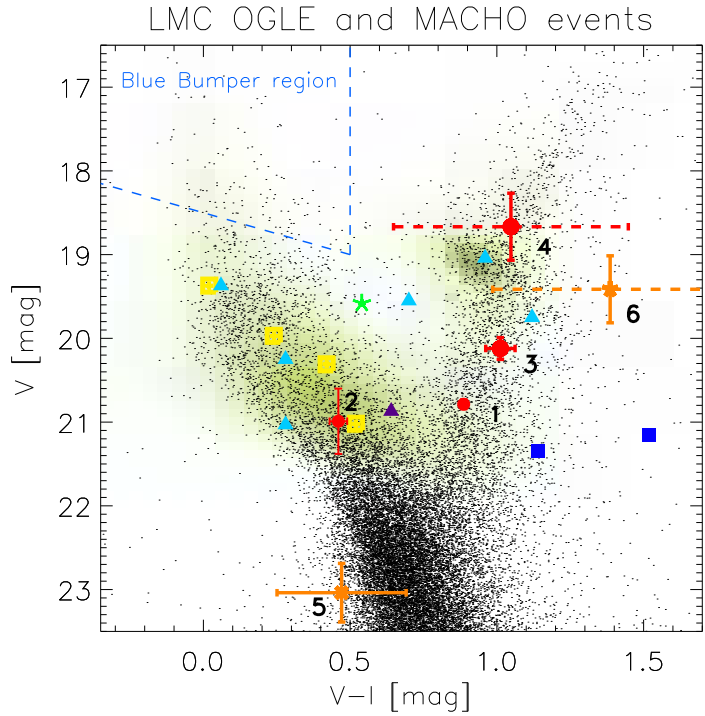}
\caption{Colour-magnitude diagram with OGLE (green) and HST (black) stars of the central part of the LMC. 
Source positions of the OGLE-II events are marked with small red dots with numbers 1 and 2. OGLE-III events found with automated pipeline are marked with big red dots with numbers 3 and 4, while the remaining two potential candidates (found visually) are marked in orange and numbers 5 and 6. 
The error-bars for $V-I$ shown with dashed lines are fixed to 0.4 mag and indicate the colour of the source can not be derived and is assumed to be equal to the colour of the baseline.
MACHO events are marked as follows: binary event \#9 (green star), candidates for self--lensing (yellow squares), confirmed thick-disk lenses \#5 and \#20 (dark blue filled squares) and remaining candidates (blue triangles). 
Event \#7 which exhibited non-microlensing variations in the OGLE-III data is shown in violet.}
\label{fig:cmd}
\end{figure*}

\begin{figure*}
\includegraphics[width=10.0cm]{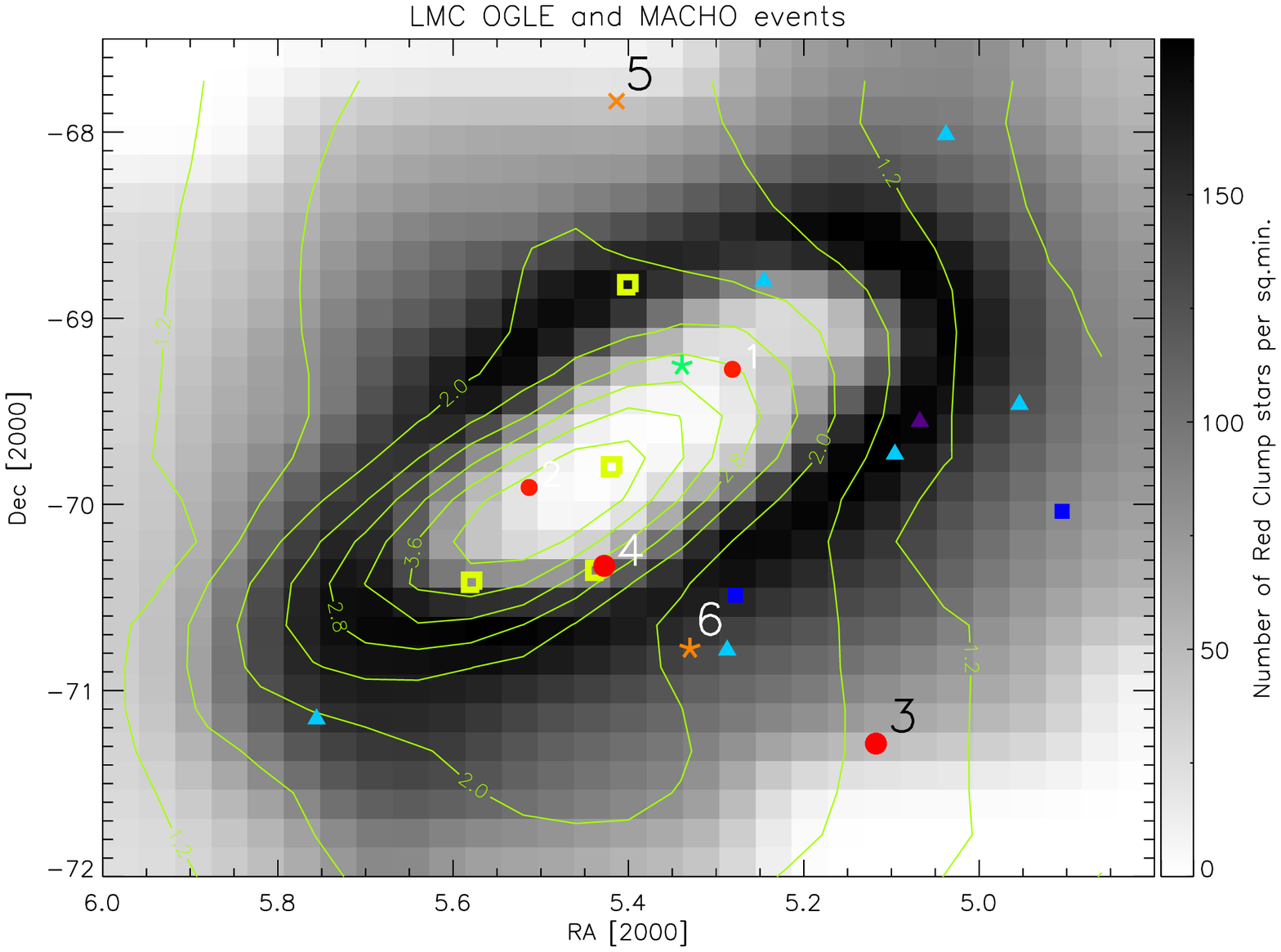}
\caption{Map of the LMC with OGLE-II (1,2), OGLE-III (3-6) and MACHO candidates for microlensing events. Colour coding of the events is the same as that in Fig. \ref{fig:cmd}.
The grey-scale background is based on the density of Red Clump stars. The contour shows the self-lensing optical depth map from \citet{Mancini2004}.}
\label{fig:mapevents}
\end{figure*}

\section{Results}
\label{sec:results}

Among the 19 million objects from 40 square degrees of the LMC observed for 8 years by the OGLE--III we found 2 firm candidates for standard microlensing events present in the All Stars and Bright Stars Samples. 
In addition, we discovered 2 more potential candidate events, which were not found by our automated search pipeline. 
One of them resembles a standard microlensing light curve, but was below our magnitude threshold, while the other is a candidate for a binary microlensing event.

Table \ref{tab:events} gathers all the information about these candidates, listing their coordinates, OGLE--III fields in which they were located, 
OGLE database star identification number, baseline magnitudes in the $I$- and $V$-band from a microlensing fit and derived magnitude and colour of the lensed source (where possible).

\begin{table*}
\caption{Microlensing events candidates detected in the OGLE--III LMC data.}
\label{tab:events}
\begin{center}
\begin{tabular}{ccccccccc}
\hline
Event's name             & RA          & Dec       & field & database      &  baseline $I$      & baseline $V$ & source $I$ & source $V-I$\\
                               & [J2000.0]   & [J2000.0] &         & star id & [mag]                   & [mag]                          & [mag] &          [mag]\\
\hline
OGLE-LMC-03          & 5:07:03.63 & -71:17:06.3 &  LMC122.1 & 15630 & 18.42    & 19.52        & 19.18            & 1.05 \\
{\scriptsize(EWS: OGLE-2007-LMC-01)} &    &     &                &       & $\pm$0.01 &$\pm$0.01  &   $\pm$0.10   & $\pm$0.05  \\
 & & & & & & & \\
OGLE-LMC-04          & 5:25:39.58 & -70:19:49.7 & LMC163.6 & 89262 & 17.33      & 18.45        &  17.65           & - \\
       &     &                        &             &                                            &$\pm$0.01 & $\pm$0.01 &  $\pm$0.02   &  \\
 & & & & & & & \\
\hline
OGLE-LMC-05          & 5:24:49.11 & -67:50:04.8 & LMC159.5 & 26848 & 21.27     & 22.31        & 22.62        & 0.51 \\
       &     &                        &             &                                          &$\pm$0.03  & $\pm$0.06 & $\pm$0.13 & $\pm$0.22 \\
 & & & & & & & \\
OGLE-LMC-06          & 5:19:47.80 & -70:46:26.6 & LMC105.4 & 25643 & 18.06     & 19.52  & - & - \\
(EROS2-LMC\#15)       &     &        &      &                                       &$\pm$0.01    & $\pm$0.01 & & \\
 & & & & & & & \\
\hline
\end{tabular}
\end{center}
\end{table*}

\begin{table*}
\caption{Parameters of the standard Paczy{\'n}ski microlensing model fits to the OGLE-III events. Only events modelled with a standard microlensing model are shown.}
\label{tab:models}
\begin{tabular}[h]{lrrrrrr}
\hline
\multicolumn{7}{c}{OGLE-LMC-03}\\
parameter & \multicolumn{2}{c}{5-parameter fit} & \multicolumn{2}{c}{4-parameter fit} & \multicolumn{2}{c}{7-parameter fit}\\
\hline
$\t0$   \dotfill        & 4136.8  & $\pm0.1$  &  4137.0 & $\pm0.1$ & 4136.8 & $\pm0.1$ \\
 & & & & & &\\
$\tE$   \dotfill        & 34.97  & $^{+4.29}_{-3.84}$ &  24.65 & $\pm0.41$ & 36.16 & $^{+4.23}_{-3.80}$ \\
 & & & & & &\\
 $\u0$  \dotfill        & 0.08972 & $^{+0.03824}_{-0.03508}$ & 0.22681 & $\pm0.00903$ & 0.08124 & $^{+0.03554}_{-0.03391}$ \\
 & & & & & &\\
$\I0$  \dotfill         & 18.423 & $\pm0.002$  & 18.373 & $\pm0.002$ & 18.423 & $\pm0.002$ 
\\
 & & & & & &\\
 $\fsi$ \dotfill        & 0.514 & $^{+0.117}_{-0.908}$   &  1.0   & --- & 0.486 & $^{+0.104}_{-0.082}$ \\
 & & & & & &\\
 $V_0$  \dotfill        & ---   & ---          & ---    & --- & 19.517 & $\pm0.006$ \\
 & & & & & &\\
 $f_{\rm S_V}$ \dotfill & ---  & ---           &  ---   & --- & 0.520 & $^{+0.114}_{-0.089}$ \\
 & & & & & &\\
 $\chi^2$\dotfill  & 1114.1 &                     &  1123.0 &   & 1309.3 &  \\
 & & & & & &\\
 ${\chi^2\over N_{dof}}$\dotfill  & 2.52  &       &  2.53 &  & 2.71 &  \\
\hline
\hline
\multicolumn{7}{c}{OGLE-LMC-04}\\
parameter & \multicolumn{2}{c}{5-parameter fit} & \multicolumn{2}{c}{4-parameter fit} & \multicolumn{2}{c}{7-parameter fit}\\
\hline
$\t0$   \dotfill        & 2227.9  & $\pm0.5$  &  2227.9 & $\pm0.5$ & --- & \\
 & & & & & &\\
$\tE$   \dotfill        & 32.76  & $^{+9.12}_{-12.57}$ &  29.26 & $\pm0.59$ & --- &  \\
 & & & & & &\\
 $\u0$  \dotfill        & 0.87763 & $^{+0.90395}_{-0.27748}$ & 1.0440 & $\pm0.00852$ & --- & \\
 & & & & & &\\
$\I0$  \dotfill         & 17.238 & $\pm0.001$  & 17.238 & $\pm0.001$ & ---&  \\
 & & & & & &\\
 $\fsi$ \dotfill        & 0.70253 & $^{+2.9765}_{-0.34558}$   &  1.0   & --- & --- & \\
 & & & & & &\\
 $V_0$  \dotfill        & ---   & ---          & ---    & --- & --- & \\
 & & & & & &\\
 $f_{\rm S_V}$ \dotfill & ---  & ---           &  ---   & --- & --- & \\
 & & & & & &\\
 $\chi^2$\dotfill  & 911.4 &                     &  911.5 &   & --- &  \\
 & & & & & &\\
 ${\chi^2\over N_{dof}}$\dotfill  & 1.56  &       &  1.56 &  & --- &  \\
\hline
\hline
\multicolumn{7}{c}{OGLE-LMC-05}\\
parameter & \multicolumn{2}{c}{5-parameter fit} & \multicolumn{2}{c}{4-parameter fit} & \multicolumn{2}{c}{7-parameter fit}\\
\hline
$\t0$   \dotfill        & 3106.0  & $\pm3.9$  &  3108.6 & $\pm4.1$ & 3105.6 & $\pm$3.8\\
 & & & & & &\\
$\tE$   \dotfill        & 224.1  & $^{+198.1}_{-86.1}$ &  135.6 & $\pm12.1$ & 346.9 &  $^{+313.5}_{-128.8}$\\
 & & & & & &\\
 $\u0$  \dotfill        & 0.12014 & $^{+0.11878}_{-0.065249}$ & 0.24846 & $\pm0.01389$ & 0.070525 & $^{+0.058690}_{-0.037110}$    \\
 & & & & & &\\
$\I0$  \dotfill         & 21.25 & $\pm0.03$  & 21.24 & $\pm0.03$ & 21.27 & $\pm$0.03 \\
 & & & & & &\\
 $\fsi$ \dotfill        & 0.43664 & $^{+0.51432}_{-0.24285}$   &  1.0   & --- & 0.25126 & $_{-0.13285}^{+0.22615}$\\
 & & & & & &\\
 $V_0$  \dotfill        & ---   & ---          & ---    & --- & 22.31 & $\pm$0.06 \\
 & & & & & &\\
 $f_{\rm S_V}$ \dotfill & ---  & ---           &  ---   & --- & 0.46176 & $_{-0.26435}^{+0.50423}$ \\
 & & & & & &\\
 $\chi^2$\dotfill  & 274.1 &                     &  275.2 &   & 330.46 &  \\
 & & & & & &\\
 ${\chi^2\over N_{dof}}$\dotfill  & 0.673  &       &  0.674 &  & 0.749 &  \\
\hline
\end{tabular}

\end{table*}

\begin{figure*}
\center
\includegraphics[width=7cm]{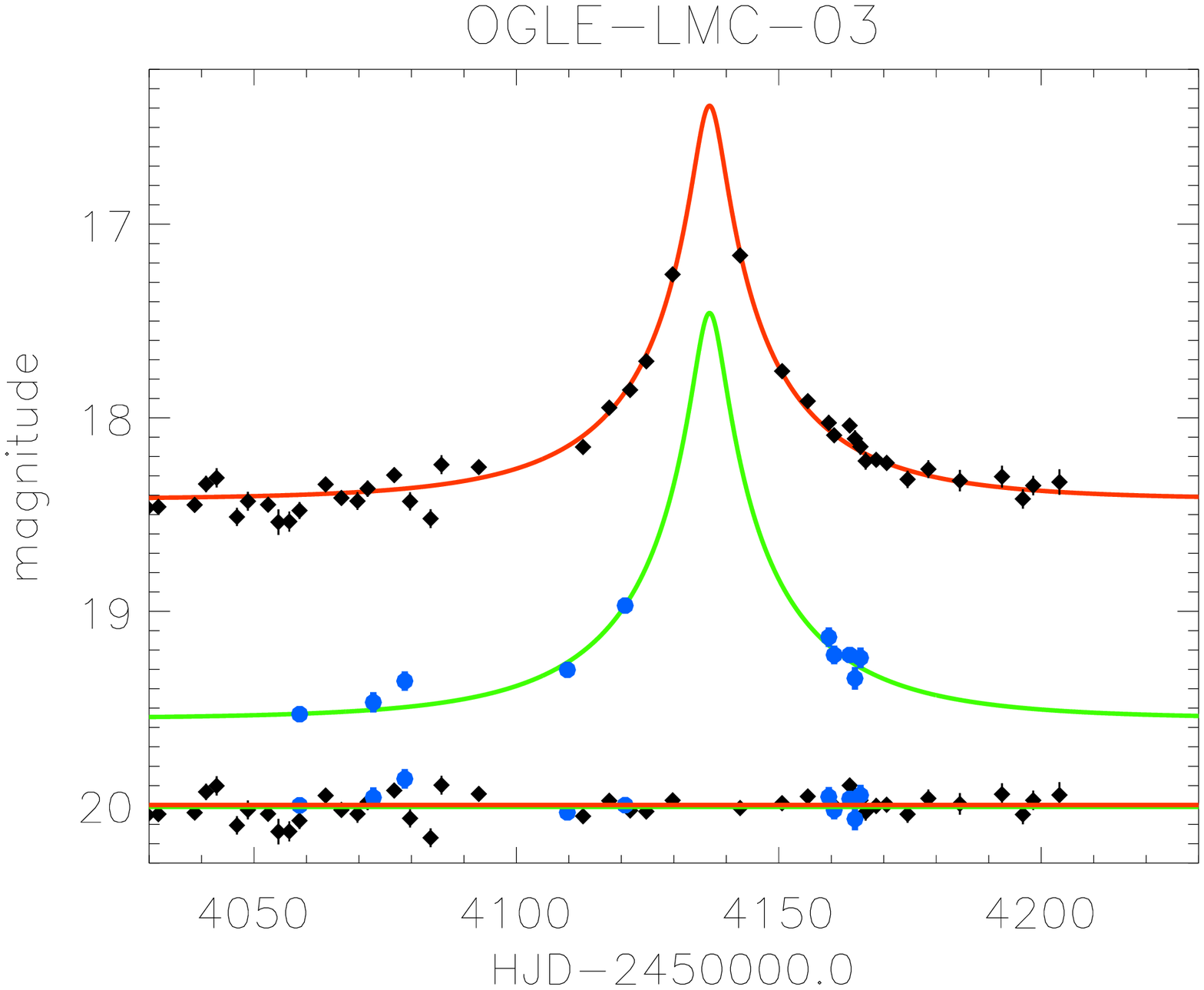}
\includegraphics[width=7cm]{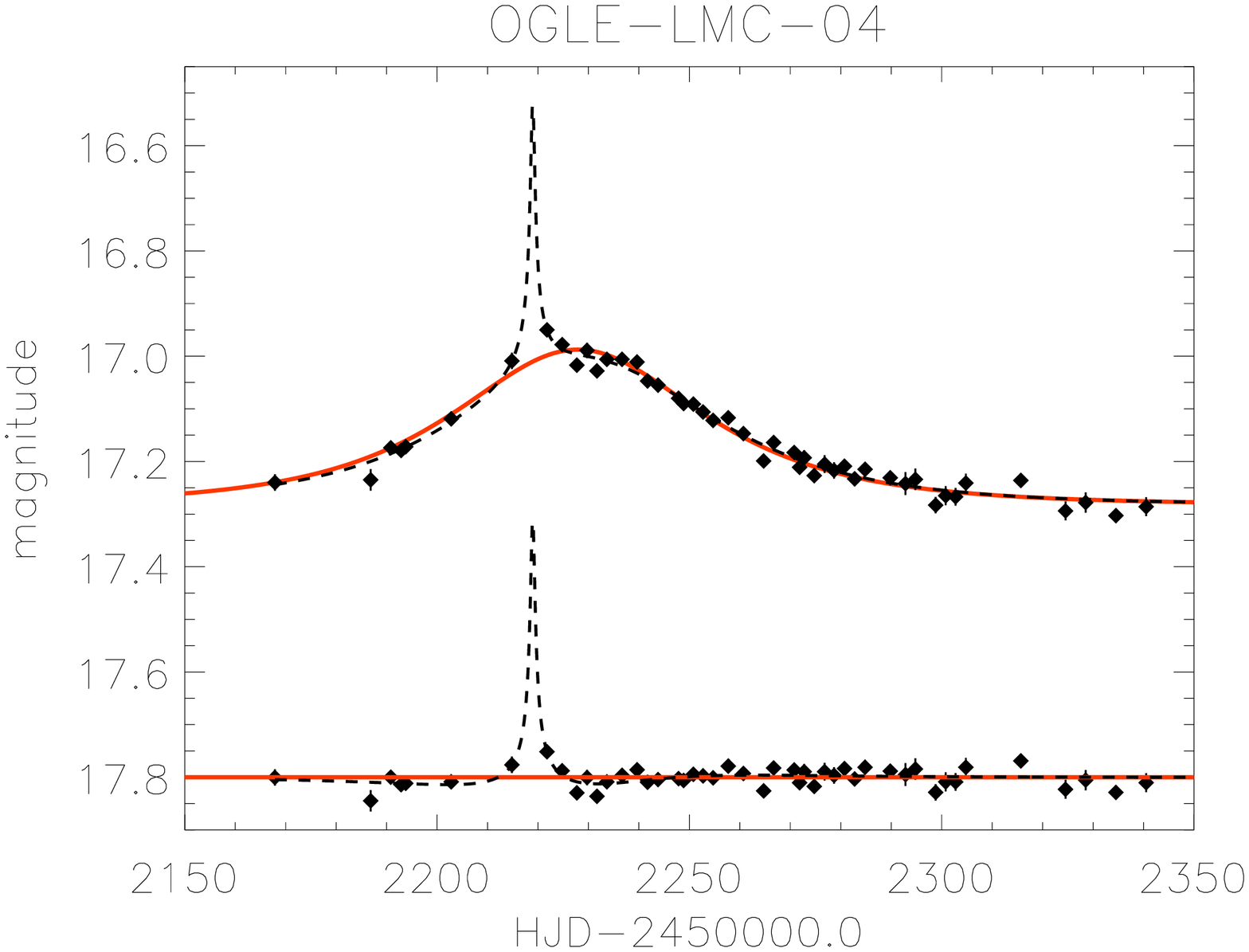}\\
\includegraphics[width=7cm]{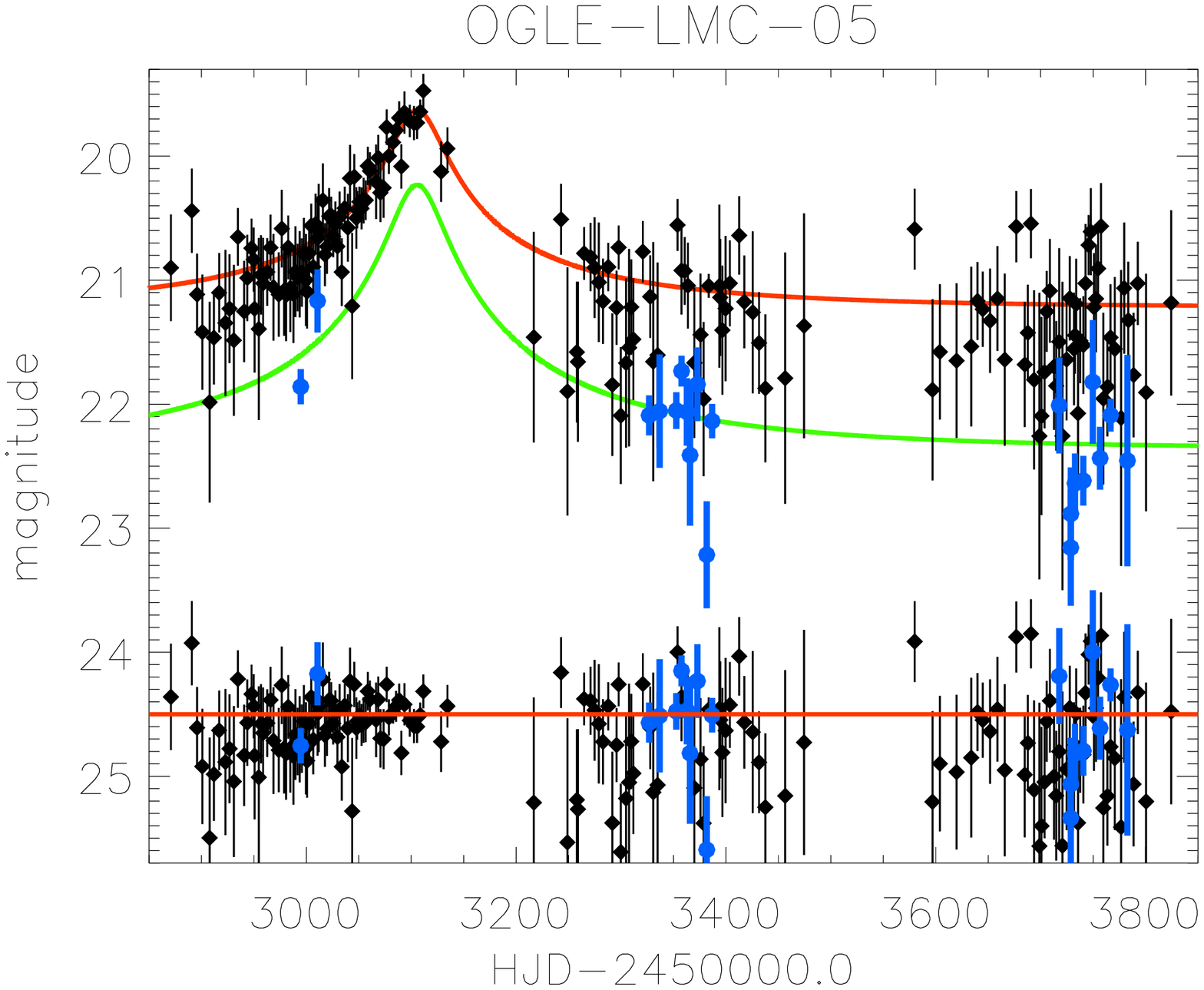}
\includegraphics[width=7cm]{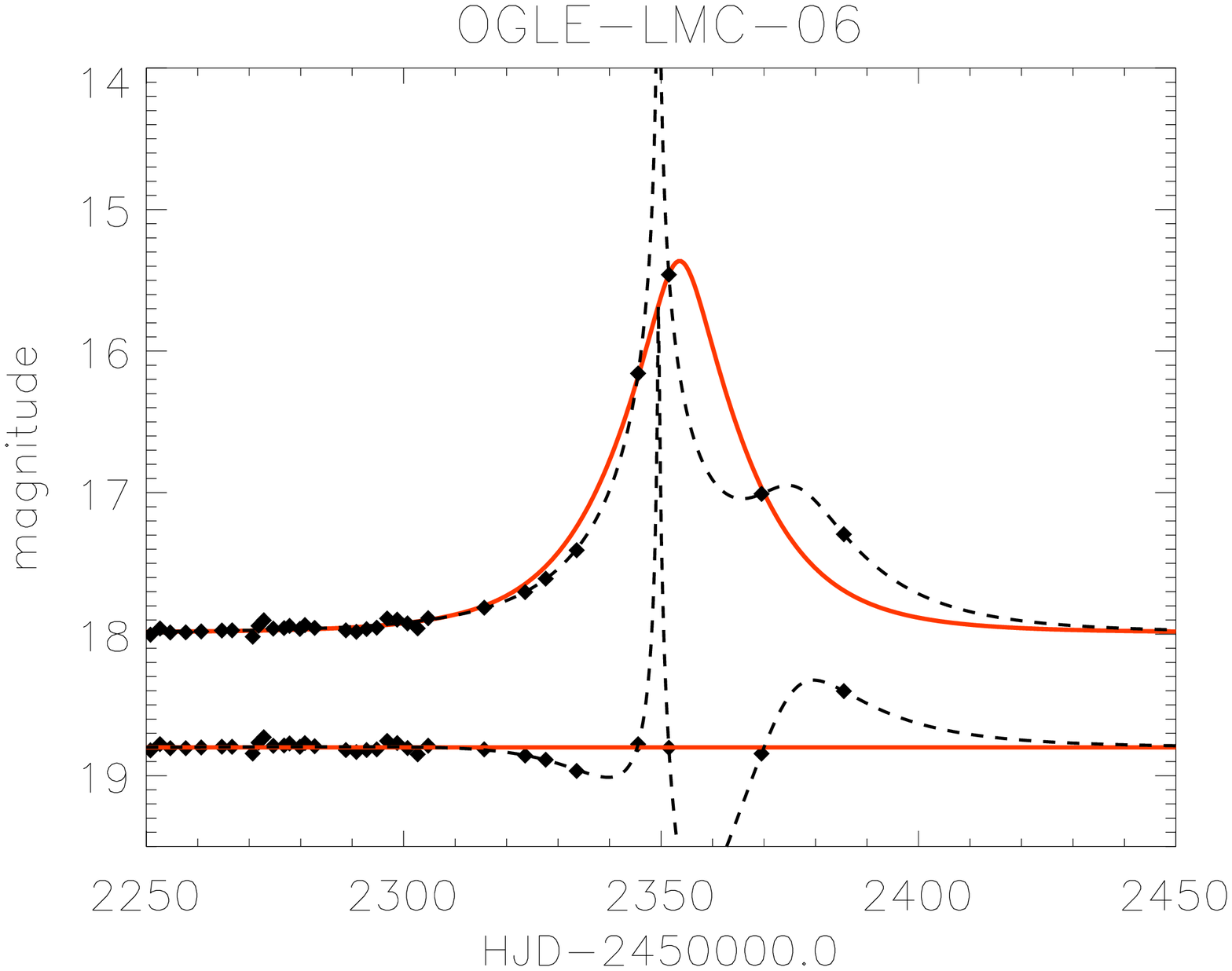}
\caption{Light curves and microlensing models of candidates for microlensing events detected in the OGLE--III LMC data. The best-fitting standard model is shown with solid lines and the best-fitting binary source model is shown with dashed lines. 
Residuals for each model are shown at the bottom of each plot.
The red curve and black dots show $I$-band data and the green curve and blue dots show $V$-band data, where available. Events OGLE-LMC-03 and OGLE-LMC-04 were found with the automated pipeline, whereas the remaining events were found by visual inspection of the data set.}
\label{fig:events}
\end{figure*}

Light curves of the events are shown in Fig. \ref{fig:events}, along with the best microlensing models obtained for the available $I$- and $V$-band OGLE data. 
Parameters of the standard Paczy{\'n}ski model fit for applicable events are gathered in Table \ref{tab:models}. Models with 7 parameters were performed on $I$ and $V$ data, where available. 5- and 4-parameters models were fit to the $I$-band data only, with blending parameter as a free parameter and fixed to unity, respectively.

Fig. \ref{fig:cmd} shows derived positions of the sources of all OGLE--III candidates for microlensing events on the colour-magnitude diagram (CMD), with the locus of LMC stars based on the OGLE and HST data. 
Also plotted are the positions of two OGLE--II events from Paper I and events reported by MACHO, colour- and shape-coded to differentiate between self--lensing candidates according to \citet{Mancini2004}, the binary event MACHO-LMC-9, the thick disk lens candidates MACHO-LMC-5 \citep{MACHOLMC5} and MACHO-LMC-20 \citep{MACHOLMC20} and the remaining candidates.

Fig. \ref{fig:cmdlenses} shows the same CMD, but with positions of the baselines of OGLE--III events and estimated locations of the lenses in these events.
The lens locations were derived based on an assumption that the entire blending light belongs to the lens. 

Fig. \ref{fig:mapevents} again presents all events ever detected towards the LMC from MACHO, OGLE--II and OGLE--III, over-plotted on top of the density map of the Red Clump stars from the LMC.

Below we discuss each of the OGLE--III events.

\begin{figure}
\center
\includegraphics[width=3.5cm]{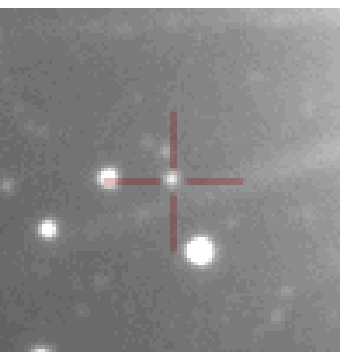}
\includegraphics[width=3.5cm]{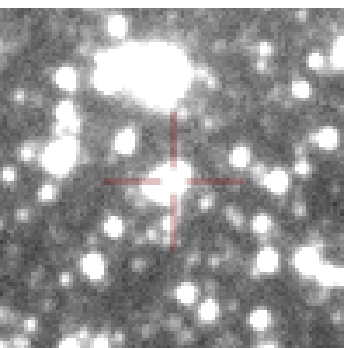}\\
\includegraphics[width=3.5cm]{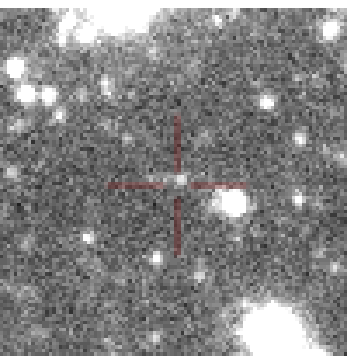}
\includegraphics[width=3.5cm]{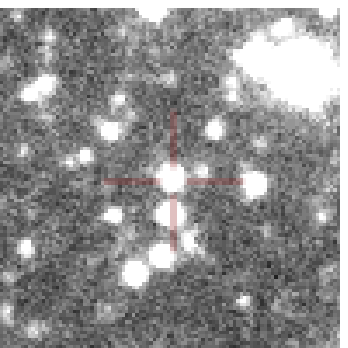}
\caption{Finding charts of the four candidates for microlensing events. East is to the right, North is down. The side of each chart is 26 arc seconds. Events 03 and 04 are shown on the top (from left to right, respectively), events 05 and 06 are shown at the bottom. A cross marks the object on which the microlensing brightening was detected. }
\label{fig:charts}
\end{figure}

\subsection{OGLE-LMC-03}
This event was the only event alerted by the OGLE's Early Warning System towards the LMC and was designated as 2007-LMC-01\footnote{http://ogle.astrouw.edu.pl/ogle3/ews/2007/lmc-001.html}.

The baseline of this candidate exhibits some irregular variations. These are the most likely caused by the presence of the very bright (6 mag, HD 33923) star (saturated on OGLE frames) less than one arc minute away.
Otherwise, the baseline does not show any additional event-like phenomenon, supported by the MACHO and EROS observations since 1993. 
This makes a very strong case for the microlensing origin of this event.

A single lens microlensing model fit to the data was performed using $I$- and $V$-band OGLE data and its parameters are gathered in Table \ref{tab:events}.
The event's light curve along with the best fit model is shown in Fig. \ref{fig:events}.

From the multi-band fit we were able to derive the source magnitude and colour of $\IS=19.18\pm0.10$ mag and $\VIS=1.05\pm0.05$ mag.
This places the source on the red giant branch of the CMD (Fig. \ref{fig:cmd}), indicating the source intrinsically belongs to the LMC.

Field LMC122, in which OGLE-LMC-03 is located, is one of the sparsest according to its stellar density (see Table \ref{tab:fields}). 
At that density and magnitude level, there is about 9 per cent of stars being blended with another one (see Section \ref{sec:blending}), therefore we can safely assume the entire remaining light in the blend constituting the baseline of the event comes from a single star - the lens. 
Another constraint comes from the astrometric measurements of the DIA residuals around the peak of the event. 
They clearly indicate lack of any shift of the centroid during the microlensing magnification, which suggests the additional flux sits exactly on top of the source star.

The blending parameters in both pass-bands obtained in the light curve fitting indicate the lens is very similar to the source in terms of the brightness and colour.
Hence, the lens may also be an LMC red giant (see Fig.\ref{fig:cmdlenses}).

\subsection{OGLE-LMC-04}
This relatively bright red star (17.3 mag) shows an event of small amplitude ($\sim$0.5 mag) just at the beginning of the OGLE--III observations in Nov 2001 (the peak occurred on $HJD=2\;452\;227.9$).
Its light curve is shown in Fig. \ref{fig:events}.

There was no $V$-band OGLE data available during the course of the event, but 37 $V$-band data points after the event allowed a derivation of the colour of the blend. A standard microlensing model fit to the $I$-band data shown in Fig. \ref{fig:events} (solid line) is best described with the time-scale of $\te=32.76^{+9.12}_{-12.57}$ days, a maximum amplification of $A=1.45$ and the blending parameter $\fsi=0.70^{+2.98}_{-0.35}$. 
Blending less than 1 indicates there is an additional light present in the overall flux of the object. 
The source magnitude can be calculated from the model as $\IS=17.65\pm0.02$ mag, but the colour can be only assumed to be equal to the blend's, \ie $(V-I)_{\rm S}\sim 1.1$. 
Even with these values and their uncertainties, the CMD location of the source can be linked to the LMC stars locus at the red giant branch, just above the Red Clump (see Fig. \ref{fig:cmd}). 
If all additional light within a blend comes from the lens (also hinted by zero-shift of the light centroid) its brightness can be calculated, however because there were no $V$-band data during the event we could not derive its colour. The estimated location of the lens in this scenario is shown on Fig. \ref{fig:cmdlenses}.

We note the large positive error on the blending parameter, which might indicate that the standard microlensing model is in fact not favoured for this event.
Its relatively bright baseline and small amplification may also suggest the brightness bump is caused by some kind of a variable star, for example a ``Blue Bumper'' blended with a red star to move its colour towards the red part of the CMD.
Moreover, the light curve indicates some small asymmetry around the peak, which can be reproduced with a microlensed binary source (dashed line in Fig.\ref{fig:events}) with the goodness of fit $\chi/N_\mathrm{dof}=1.52$. However, that model has very little constraints and the ``wiggle'' around the peak could be as well caused by instrumental inaccuracies.
Further photometric observations of this star are necessary as ``Blue Bumpers'' usually have secondary brightenings after a decade or so.

\subsection{OGLE-LMC-05}

This event was not detected in our regular search procedure described above, because its baseline magnitude (21.22 mag) was well below our threshold (20.4 mag). 
It was found when the search for events was conducted with magnitude cut removed (see Table \ref{tab:conditions}), which means it would pass through the pipeline if the magnitude cut was different.

The microlensing model fit to its $I$- and $V$-band light curves (shown in Fig. \ref{fig:events}) yielded the time-scale of $\te=347\pm178$ days and $\u0=0.0705\pm0.0447$ with blending parameters of $\fsi=0.2512\pm0.1643$ and $\fsv=0.4618\pm0.3460$.
The goodness of fit to multi-band data was $\chi^2/dof=0.749$, whereas when only $I$-band data were used it was $\chi^2/dof= 0.676$.
The single passband model returned also a relatively long time-scale of $\te=224\pm115$ days with $\u0=0.1201\pm0.0820$ and $\fsi=0.4366\pm0.3201$.

From the blended multi-colour models we were able to estimate the magnitude and colour of the source as $\IS=22.62\pm0.13$ mag and $\VIS=0.51\pm0.22$, which locates the source on the blue end of the main sequence of the LMC (see Fig. \ref{fig:cmd}).

\begin{figure}
\includegraphics[width=8.0cm]{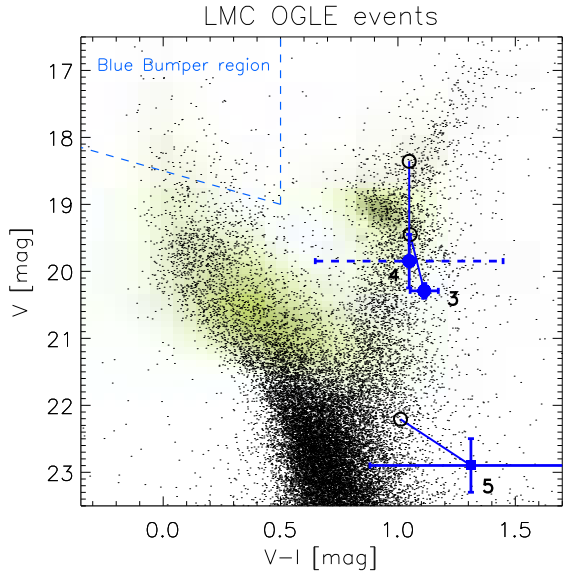}
\caption{Colour-magnitude diagram as in Fig. \ref{fig:cmd} showing estimated positions of the lenses of the candidates for microlensing events found in the OGLE-III data towards the LMC.
Open circles mark the positions of the blended objects as seen at the baseline of each event. 
The error-bars for $V-I$ of the lens shown with dashed lines indicate that the colour of the lens could not be determined and was taken from the blend/baseline.
}
\label{fig:cmdlenses}
 \end{figure}

Severe blending indicates there is a lot of additional light within the seeing disk of the object with the event. 
Judging from the sparse stellar density of the field LMC159 containing the event, we could assume all the remaining light belongs to the lens. 
Astrometry of the centroid from the DIA indicates no shift compared to the baseline template position. This also supports the assumption that there is no additional light taking part in the event except the source and the lens.
This places the lens at $V_{\rm L}=23.00\pm0.40$ mag and $(V-I)_{\rm L}=1.35\pm0.43$, which is far from the main LMC locus (see Fig. \ref{fig:cmdlenses}).
Its CMD location in the vicinity of the two MACHO events caused by thick-disk lenses suggests this could be another example of a Galactic lens towards the LMC.
This, however, is relatively difficult to confirm with a follow-up using currently available instruments given the very faint magnitudes of both the source and the lens and large uncertainties of event's parameters.

\subsection{OGLE-LMC-06}

This candidate was also not detected by the regular search pipeline, but was visually found in the data when inspecting large signal to noise ratio light curves.
Its light curve exhibits asymmetric bump with amplitude of about 3 mag, but the peak is covered very sparsely by the OGLE observations. 
However, single lens model does not reproduce the light curve.
The simplest static binary source model is doing a much better job (see Fig. \ref{fig:events}), however due to insufficient number of datapoints we can not exclude numerous possible binary lens models here.
The binary source model gave time-scale of about 27 days and amplifications of 3 and 137 on each of the components with $\chi/N_\mathrm{dof}=1.51$, compared to $\chi/N_\mathrm{dof}=3.55$ for a standard single source single lens model.

The lack of a sensible model for the event prevents a derivation of the position of the lensed source on the Colour-Magnitude Diagram. 
The colour and magnitude of the blend (source and lens, plus any additional blending light), marked on Fig. \ref{fig:cmd} places it about 0.4 mag to the right of the red giant branch of the LMC. 
If this was the position of the source, it would suggest it does not belong to the LMC and is either in the foreground (in the Milky Way halo) or in the background behind the LMC. 

This candidate was already flagged by the EROS group \citep{TisserandEROSLMC} as anomalous. Their data also indicate a slight reddening during the brightening of the event. This could hint towards some kind of nova eruption, however the light curve in its rising part is too smooth for a typical outburst-like object.
Taking all this into account we can not firmly conclude on the nature of this event.

\subsection{Historic events}
For completeness, we checked the previously known microlensing event candidates if their OGLE-III counterparts were exhibiting additional variation atypical for microlensing events.
It was not necessary for the OGLE-II events as their light curves collected during the OGLE-III were already investigated in Paper I, where it was confirmed they had constant baselines in years 2001-2009.

MACHO group \citep{AlcockMACHOLMC} have reported 17 candidate events and all of them were located within the OGLE-III fields. 
We confirm the constant baseline behaviour during the OGLE-III phase in most of them, except MACHO-LMC-7 and MACHO-LMC-23. 
The secondary peak in event MACHO-LMC-23 we see in our data occurred at JD$\approx2\;452\;250$ and was already reported by \citet{TisserandEROSLMC}.
This object was on our list of objects which were subject to cuts applied above, but its light curve was clearly asymmetrical and it was removed in the cut 8 with its blended model not converging and with large negative blending parameter.

Event MACHO-LMC-7 originally passed through improved selection criteria for MACHO events and was included in the optical depth determination of \citet{BennettMACHOLMC}. 
The OGLE-III light curve of this object reveals three small amplitude bumps separated by $\sim$2 years, all resembling microlensing curve with time-scales between 20 and 40 days and amplitudes around 0.5 mag.
It clearly indicates this is not a genuine microlensing event but some sort of repeating outbursting variable star.
It contributed about 10 per cent to the optical depth derived in \citet{BennettMACHOLMC}, therefore their value should be now decreased to about $\tau_{\rm{LMC}}\approx 0.9 \times 10^{-7}$.

Among the events reported by EROS group \citep{TisserandEROSLMC} we successfully cross-matched with the OGLE-III data only two candidates: EROS-LMC-1 and EROS-LMC-11.
The latter showed constant baseline over the course of the third phase of the OGLE project, however the former exhibited a nice symmetric microlensing-like bump at JD$\approx2\;453\;410$ and, as mentioned above, actually passed through our search pipeline.
EROS has already informed about the second peak they detected in that event occurring after nearly 5 years after the first one.
The bump we detected happened after another 7 years, showing the underlying contaminant population of mysterious bumpers towards the LMC can exhibit a microlensing-like repeating episodes of brightening with random periodicity.
The only possibility of ruling them out is to rely on as long time-baseline of observations as possible.


\section{Blending}
\label{sec:blending}
The Earth's atmosphere blurs stellar images. Astronomical images taken with medium- and large-sized telescopes are almost never 
diffraction limited, unless active and/or adaptive optics is used. The Earth's atmosphere prevents us from seeing much sharper and 
clearer images than we would otherwise see in its absence. 
Our atmosphere affects crowded stellar fields the most, such as the Magellanic Clouds,
as many stars merge together into composite objects (``blends'') that often can be no longer resolved by PSF fitting. 
Since the gravitational microlensing is equally likely to happen on any star in a blend, no matter if it is bright or faint, we would like to know how many stars contribute to one composite object, because in general the number of observed objects is not equal, and usually less, than the real number of monitored stars. 
This is of particular importance for the optical depth estimator, since at least
three of its ingredients are directly affected by blending. The most obvious dependent quantity is the number of monitored stars.
Without correction for blending our estimate of the optical depth would be overestimated. 
Another directly dependent quantity from the optical depth estimator is the detection efficiency of microlensing events.
Detecting a microlensing event on a bright source blended with fainter stars would yield almost 100 per cent detection efficiency.
On the other hand, in case where a faint microlensed star is blended with a much brighter star the detection efficiency for such an event will be very small. 
Also the time-scale of an event is affected by blending and appears shorter with blending 
(see e.g., \citealt{Wozniakblending}; \citealt{Smith2007blending}).
It is critically important to understand the blending and its effects on the optical depth estimator. 

\begin{figure}
\includegraphics[width=8.5cm]{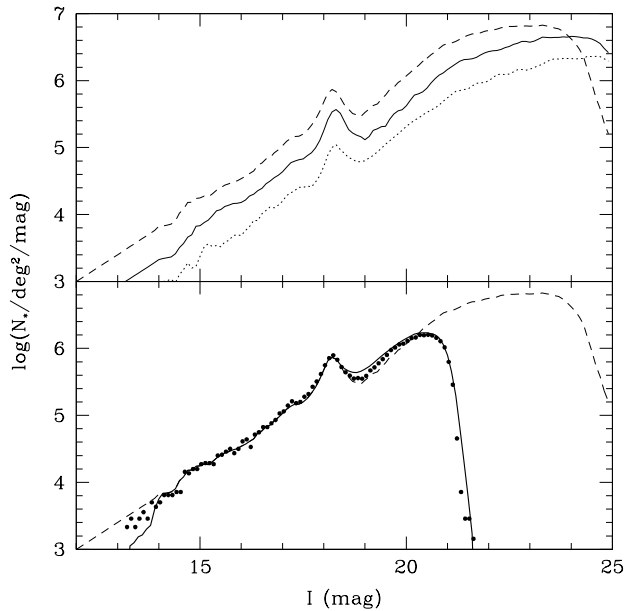}
\caption{Top panel: Three combined {\it HST} and OGLE-III luminosity functions for the LMC used in the blending simulations. 
The top line (dashed) is for field LMC162.6, middle (solid) is for LMC119.2 and the bottom (dotted) is for LMC136.1. 
Bottom panel: Example of recovered luminosity function (dots) for the field LMC162.6, closely following the observed OGLE-III function (solid line).}
\label{fig:LF}
\end{figure}

The amount of blending can be estimated by comparing the ground-based images with much more detailed archival {\it HST} images and then grouping fields according to their stellar density and applying a fixed correction to the number of monitored stars for each density level.
Such a method was applied in Paper I and Paper II for the OGLE-II data.
However, the area covered in a single {\it HST} image is tiny, hence such comparison results in very low number statistics of cross-matched stars.
Moreover, OGLE--III fields cover a very wide spectrum of stellar densities, which is difficult to cover with enough number of archival HST images.
Therefore to address the issue of blending here, we performed simulations of the OGLE-III images based on combined stellar luminosity functions derived from the {\it HST} and OGLE images.

\subsection{Simulations}
Based on known properties of the OGLE-III images, such as the point spread function's (PSF) size and shape as well as the shape of the underlying stellar luminosity function, we are able to simulate OGLE-III images with different stellar densities and luminosity functions.
We used the LMC photometry from 
{\it the HST Local Group Stellar Photometry} archive\footnote{{\tt http://ganymede.nmsu.edu/lg}} \citep{Holtzman2006}. 
Three selected HST fields, {\sc lmc\_u4b115}, {\sc lmc\_u65008}, and {\sc lmc\_u65007}, were
each observed in F555V (F814W) filter with the WFPC2 for 2000 sec (2000 sec), 2560 sec (2460 sec), and 1560 sec (1560 sec) seconds. 
These fields were calibrated to the standard $I$-band filter and were located in OGLE's fields
LMC162.6 (dense), LMC119.2 (medium density), and LMC136.1 (sparse). 
Then, the OGLE and {\it HST} luminosity functions were combined with a stitching point around the red clump $I$-band magnitude, \ie $I\approx18.2$ mag. 
Due to low number statistics for stars with $I<14$ mag we approximated the luminosity function with $\log(N)\propto 0.4\times I$.
Each of the luminosity functions had a slightly different shape shown in Fig. \ref{fig:LF},
but, as show later, this affects the correction factor for the number of monitored stars by less than 5 per cent (Fig. \ref{fig:CF2}).

\begin{figure}
\includegraphics[width=8.5cm]{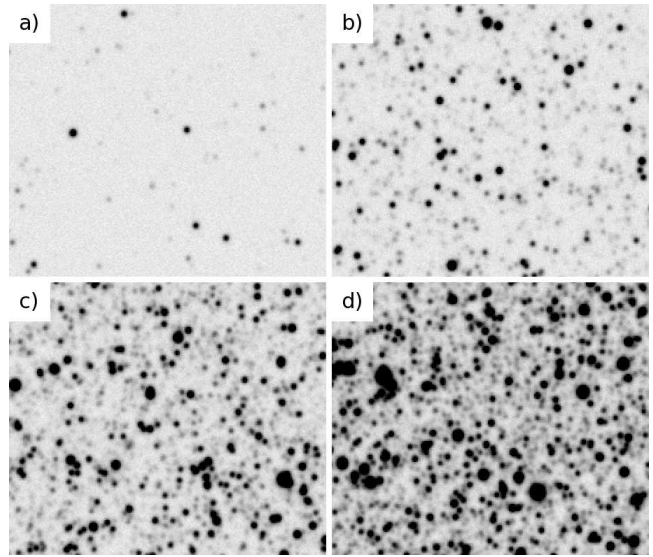}
\caption{Simulated OGLE-III LMC images. The stellar density increases from panel 
a) with $\log(N_*/\rm{CCD~chip})=3.82$, through panels
b) with $\log(N_*/\rm{CCD~chip})=4.83$, and
c) with $\log(N_*/\rm{CCD~chip})=5.17$, up to panel 
d) with $\log(N_*/\rm{CCD~chip})=5.32$.}
\label{fig:SimOGLE}
\end{figure}

Before each simulation the basic parameters of the OGLE-III template image were measured.
This included the PSF shapes and sizes, and the background level. 
First, we created a mock image ($800\times800$ pixels) with the background light and Poisson noise that
matched the original OGLE images. 
Next, we chose an {\it HST} stellar density as $200\times i^{3/2}$ stars/arcmin$^2$, 
where $i$ is the simulation number, and $i=1$ to 15. 
Then, we injected stars that were drawn from the combined
luminosity function (top panel in Fig. \ref{fig:LF}) and were brighter than $I \la 24.8$ mag.
The faint end ($I > 23.4$ mag) of the luminosity function had a minimal impact on our study (see Fig. \ref{fig:CF2}) as our limiting search for microlensing events is $I=20.4+3.0=23.4$ mag; the faint end simply serves as a background here.
Here, $I=20.4$ mag is the magnitude threshold for stars(blends) we allow in our search for microlensing events, hence stars fainter by $\Delta m=3.0$ mag are the faintest which can contribute to a combined blended objects as sources for events.

\begin{figure}
\includegraphics[width=8.5cm]{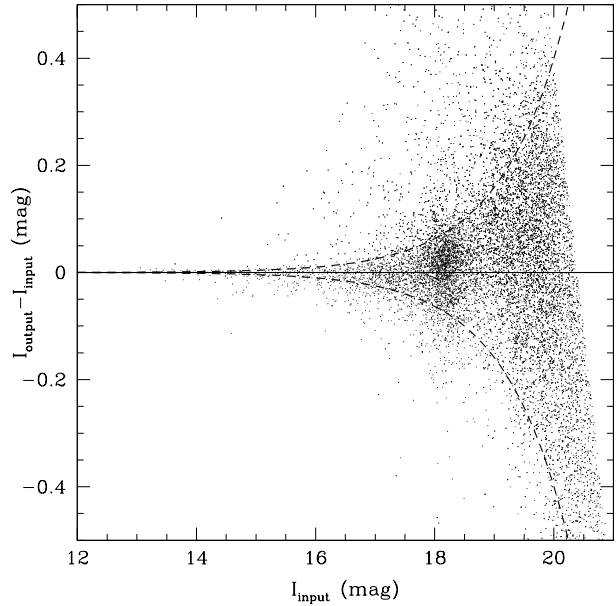}
\caption{Difference between recovered (output) and injected (input) $I$-band magnitude as a function of 
injected magnitude for a dense field with $\log(N_*/\rm{CCD~chip})\approx5.2$. 
The two dashed lines show 2$\sigma$ OGLE-III photometric uncertainties.}
\label{fig:Cal}
\end{figure}

Fig. \ref{fig:SimOGLE} shows four examples of simulated OGLE images with different stellar densities.
Once the image was created we obtained the stellar photometry in the way identical to the OGLE-III treatment of the template images using {\sc DoPhot} photometry package \citep{Dophot}. 
Finally, we matched the input (simulated {\it HST}) and output (simulated OGLE) catalogues of stars. 
After some experimentation, we find the matching radius
of $r=0.8$ OGLE-III pixels for stars with $I\ge20.4$ mag and $r=\sqrt{0.8^2 + (0.8G)^2}$ pixels, 
where $G$ is a size of the Gaussian function at the level of $I=20.4$ mag, $G=\sqrt{0.8\sigma\ln{10}(20.4-I))}$, for stars $I<20.4$ mag and PSF size $\sigma$.
The output catalogue was calibrated in such a way, that magnitudes of recovered single (not blended) stars were nearly identical to the injected ones,
but also the number of flux counts on a simulated image for a given magnitude was nearly identical to real images.
Fig. \ref{fig:Cal} shows a difference between recovered and injected magnitudes
as a function of the injected magnitude.

\begin{figure}
\includegraphics[width=8.5cm]{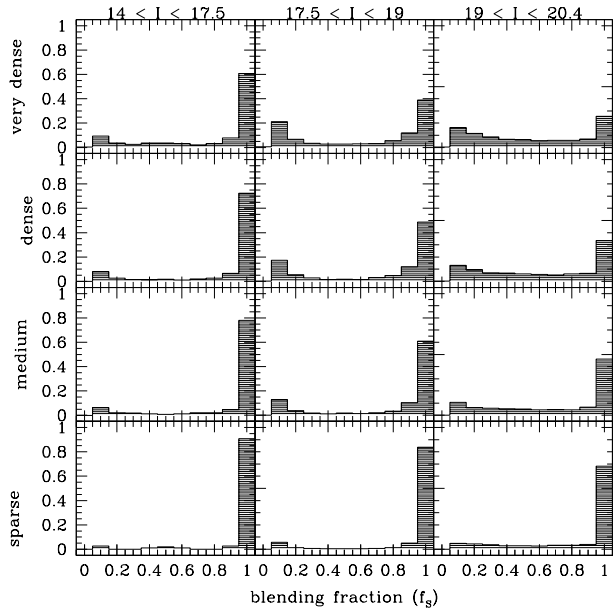}
\caption{Blending distributions for selected levels of stellar density for the simulated OGLE--III images obtained using archival HST images.
The distributions were derived in three magnitude bins, described at the top. 
The density levels shown correspond to stellar densities of $\log(N_*/\rm{CCD~chip})=(5.32, 5.24, 5.12, 4.8)$ for very dense, dense, medium and sparse, respectively. 
}
\label{fig:blending}
\end{figure}

For each OGLE star of magnitude $I_{\rm OGLE}$ within its radius $r$ we count all {\it HST} stars with brightness of $I_{\rm HST} \le I_{\rm OGLE}+3.0$ mag. 
Then, we constructed distributions of blending, where blending for each star is calculated as 
$f_{\rm star}={F_{\rm star} \over F_{\rm all~stars~in~the~blend}}$. 
The blending distributions for selected simulated density levels are shown in Fig. \ref{fig:blending}.

\subsection{Estimating the real number of monitored stars}

\begin{figure}
\includegraphics[width=8.5cm]{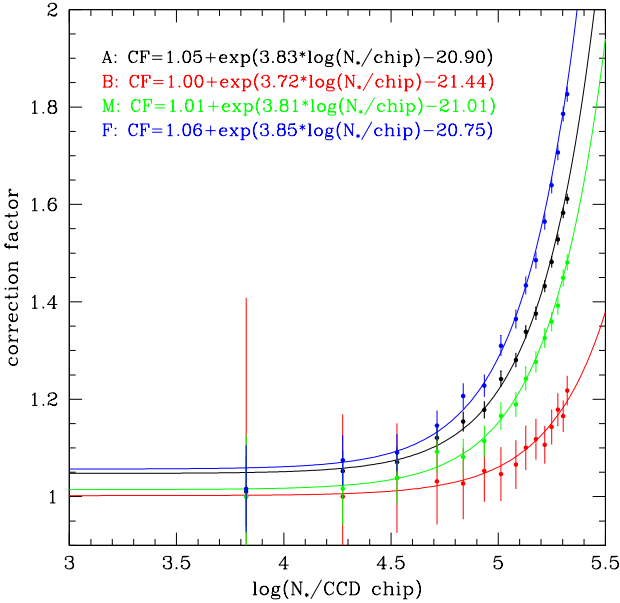}
\caption{Correction factor as a function of the number of template objects on an OGLE-III CCD chip for different magnitude bins.
The magnitude range $I\le17.5$ mag is described by function B (bright), $17.5\le I \le19.0$ mag by function M (middle), and the $19.0<I \le 20.4$ mag by function F (faint). 
The function based on the entire magnitude range, $I\le20.4$ is marked as A (all).}
\label{fig:CF}
\end{figure}

\begin{figure}
\includegraphics[width=8.5cm]{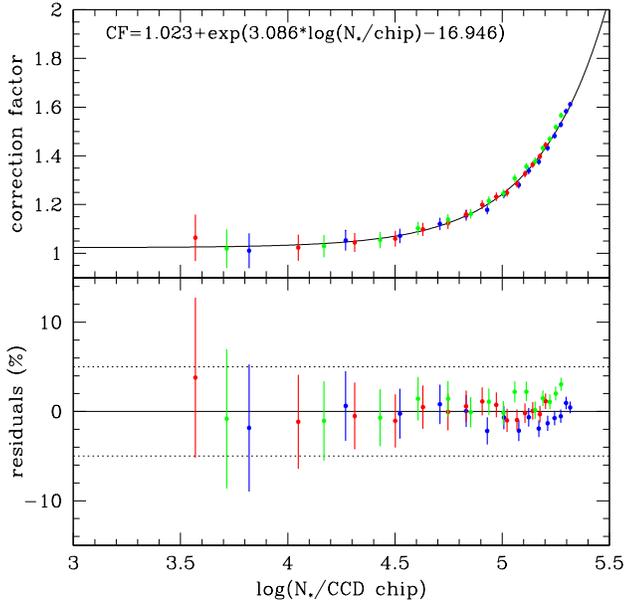}
\caption{Correction factor as a function of the number of template objects on an OGLE-III CCD chip for the entire magnitude range but for different areas of the LMC.
Blue dots are for field LMC162.6 (central part of the LMC bar), red for LMC119.2 (edge of the LMC bar), and green is for LMC136.1 (LMC's disk).}
\label{fig:CF2}
\end{figure}

By comparing number of input {\it HST} stars and resulting number of OGLE objects on simulated frames we were able to obtain a ``correction factor'' (CF). 
The correction factor converts from the number of observed stars in an OGLE-III CCD chip to the number of real stars hidden (unresolvable) due to blending, which may act as a source in a microlensing event.
Figs. \ref{fig:CF} and \ref{fig:CF2} show the CF as a function of the number of observed stars in a single OGLE-III CCD chip. 
Plot in Fig. \ref{fig:CF2} indicates that the correction factor only weakly changes (with 5 per cent) when different shapes of the luminosity functions are used (see Fig. \ref{fig:LF}).

We derived the number of stars correction for three magnitude bins, corresponding to the bins in which the blending distributions were derived:
bright $I\le17.5$ mag (B), middle $17.5\le I \le19.0$ mag (M), and faint $19.0<I \le 20.4$ mag (F).
Then, for each OGLE template object brighter than $20.4$ mag we calculated a number of corresponding unresolved stars, using the CF for the given field's density, encoded in a number of all objects found on a template image (see Table \ref{tab:fields}).
The resulting estimated number of real stars in each field (combined for all 8 CCD chips) is provided in Table \ref{tab:fields}.

In total there were around 19.4 million objects on the OGLE template images. 
These were estimated to be composed of about 22.7 million real stars, all of which could potentially be microlensed.
In the Bright Star Sample there were 5.8 million objects on the templates and 6.3 million estimated real stars.

\section{Optical depth}

\begin{figure}
\includegraphics[width=8.5cm]{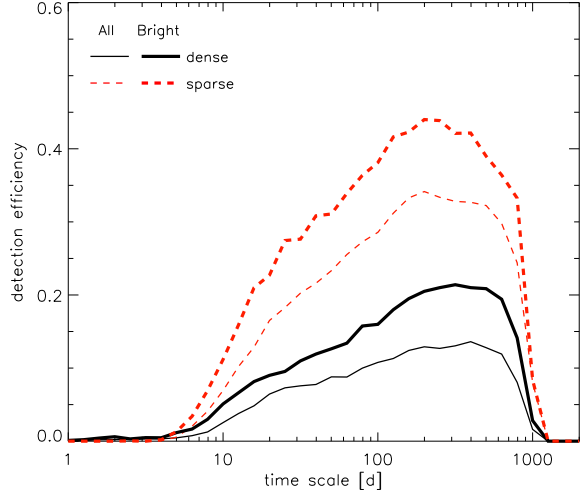}
\caption{
Microlensing events detection efficiency for OGLE-III LMC data for All and Bright Stars Samples. The efficiency strongly depends on the stellar density of the field. Curves are shown for dense (LMC163) and sparse (LMC122) fields.
}
\label{fig:eff}
\end{figure}

Above we reported a detection of 4 candidates for microlensing events.
However, in order to calculate the optical depth towards the LMC we could only use events which were returned by our automated search pipeline, \ie events which satisfied well specified criteria and were possible to describe with a point-source-point-lens microlensing model. 
These were OGLE-LMC-03 and OGLE-LMC-04. 
The remaining events were found manually and were either below the magnitude threshold (OGLE-LMC-05) or clearly not due to a single lens (OGLE-LMC-06).

The microlensing model of \citet{Paczynski1996} gives an amplification $A$ described with Eq.\ref{eq:A}.
The time-scale (Einstein radius cross time) is the only parameter of the model which is related to some physical values, but is a degenerate function of the lens mass, distance, and relative velocity.

For a number of microlensing events with time-scales ${\tE}_{i}$ and detection efficiency $\epsilon({\tE}_i)$, the optical depth towards the LMC from the OGLE-III data is calculated from the formula:
 
\begin{equation}
\label{eq:tau}
\tau_{\rm LMC-OIII}={{\pi}\over {2 N_* T_{\rm obs}}}\displaystyle \sum_i^{N_{\rm ev}} {{\tE}_{i} \over \epsilon({\tE}_i)}
\end{equation}
where $T_{\rm obs}$ is the time-span of all observations,
$N_*$ is the total number of monitored stars, 
$N_{\rm ev}$ is the total number of events
.

The detection efficiency was derived separately for each field containing a microlensing event in the Monte Carlo simulations of the events covering the range of time-scales between 1 and 1000 days. 
Figure \ref{fig:eff} shows exemplary curves of the efficiency for All and Bright Stars Samples and for dense and sparse fields. 
In these simulations all microlensing model's parameters except $\tE$ were randomly drawn from appropriate ranges: $\t0$ from $2\;452\;115$ to $2\;454\;965$ days (the entire range of the OGLE--III LMC coverage), $\u0$ from 0 to 1, $\I0$ from the luminosity function of a given simulated field and $\fs$ from blending distribution derived for that field (see Section \ref{sec:blending} and Fig.\ref{fig:blending}).

The optical depth was then calculated for $N_{\rm ev}=2$ events and their time-scales, with $N_*=22,740,000$ and $T_{\rm obs}=2850$ days.
Following \citep{TisserandEROSLMC} and Papers I and II we applied a correction for binary events for which our automated search pipeline is not sensitive to. 
It meant all the efficiencies were reduced down by 10 per cent. 

Table \ref{tab:tau} gathers the values of time-scales and detection efficiencies used for deriving the optical depth.
For both events we found the optical depth of $\tau=(0.15 \pm 0.10) \times 10^{-7}$ and $\tau=(0.16 \pm 0.12) \times 10^{-7}$ for binary-corrected efficiency.

\begin{table}
\caption{The optical depth for the two OGLE--III events found by the automated search pipeline. The columns show event name, its time-scale, detection efficiency and individual event's contribution to the overall optical depth.}
\label{tab:tau}
\begin{center}
\begin{tabular}[h]{cccc}
\hline
event & $\te$ & $\epsilon(\te)$ &$\tau_i \times 10^{-7}$ \\
\hline
\multicolumn{4}{c}{efficiency not corrected for binary events} \\
 & & & \\
OGLE-LMC-03 & 34.97$^{+4.29}_{-3.84}$  & 0.202489 & 0.04 \\
 & & & \\
OGLE-LMC-04 & 32.76$^{+9.12}_{-12.57}$ & 0.075691 & 0.11 \\
\hline
total $\tau_{\rm LMC-OIII}$   &              &          &   $0.15\pm0.10$ \\
\hline
\hline
\multicolumn{4}{c}{efficiency corrected for binary events} \\
 & & & \\
OGLE-LMC-03 & 34.97$^{+4.29}_{-3.84}$ & 0.182240 & 0.05 \\
 & & & \\
OGLE-LMC-04 & 32.76$^{+9.12}_{-12.57}$ & 0.068122 & 0.11 \\
\hline
total $\tau_{\rm LMC-OIII}$   &              &          &   $0.16\pm0.12$ \\
\hline
\end{tabular}
\end{center}
\end{table}

If we allowed for non-single events contributing to the optical depth, hence included the candidate OGLE-LMC-06 with its time-scale of around $\te=27$ days and estimated efficiency of $\epsilon(\te)=0.182767$, it would add around $0.04\times 10^{-7}$ to the total value of $\tau$.
In the other scenario, we could also speculate that OGLE-LMC-04 is actually a ``Blue Bumper'' and has to be excluded from the $\tau$ measurements.
That would reduce the value of the optical depth to $\tau=0.05\pm 0.05\times 10^{-7}$ or $\tau=\sim 0.08 \times 10^{-7}$ if OGLE-LMC-06 was included as above.

In the Bright Star Sample there were the same two events as in the All Stars Sample. 
The efficiencies of their detection were about 1.4 times larger than efficiencies derived in the All Stars Sample. On the other hand, the number of monitored stars was smaller by a factor of around 3. 
These two factors do not cancel out and the resulting optical depth for the Bright Star Sample is $\tau=0.41\pm0.29\times 10^{-7}$, which is around 2 times larger than for All Stars. 
However, given very low number of events and resulting large error bars on $\tau$, comparison between these two values can not lead to any meaningful conclusions.

\section{Discussion}

The number of microlensing event candidates detected towards the LMC during the 8 years of the OGLE-III survey is remarkably small. 
The overall optical depth calculated using the two most reliable events is
$\tau_{\rm LMC-OIII}=(0.16\pm0.12)\times 10^{-7}$ and is  smaller than obtained with two events found in the OGLE--II data ($\tau_{\rm LMC-O2}=(0.43\pm0.33)\times 10^{-7}$, Paper I). 
On the other hand, large systematic errors in both of these measurements only prove we are dealing with a very low statistics in terms of the numbers of events.
This is in clear contradiction to a significantly larger number of events claimed by the MACHO collaboration (\citealt{AlcockMACHOLMC}, \citealt{BennettMACHOLMC}), but is in agreement with results obtained with the EROS group data \citep{TisserandEROSLMC}.

Can OGLE events still be caused by the hypothetical MACHOs? 
According to the halo model of \cite{AlcockMACHOLMC} (model ``S''), if the Galaxy halo was entirely filled with dark matter compact objects it would produce the optical depth of $\tau_{\rm MACHO}=4.7\times 10^{-7}$. 
Because microlensing optical depth depends on the total mass of the lensing objects, the fraction of measured $\tau$ in $\tau_{\rm MACHO}$ gives directly a fraction of halo mass in compact objects.
Based on our measurement of the optical depth, if both microlensing events were due to MACHOs this fraction would be $f = 3\pm2$ per cent, with a typical mass around 0.2 $\msun$.
This mass can be estimated from the mean time-scale of the events found using this formula: $\log M= 2\log (<\tE> / 70)$ (\citealt{AlcockMACHOLMC}, \citealt{TisserandEROSLMC}).
The derived mass fraction is considerably smaller than $f=20$ per cent claimed by the MACHO group.

Because of the complexity of the microlensing events in general we are not able to definitively tell where the lenses and sources come from and what they are. This is only possible in case of events exhibiting additional effects, \eg parallax or when the lens is a binary, or, in some special cases, after detailed high resolution imaging and spectroscopic follow-up.
Therefore, in case of our events we can only speculate about the nature of the events; there are, however, certain hints towards a non-dark matter scenario.

Due to some line-of-sight depth of the LMC it is possible that stars of the Cloud get microlensed by other LMC's stars located in front. 
The effect of ``self-lensing'' (SL) was estimated based on the internal structure studies of the LMC by \cite{Mancini2004} and is the strongest for the LMC bar (see Fig. \ref{fig:mapevents}). 
Its contribution to the total optical depth is around $\tau_{\rm SL}=0.4\times 10^{-7}$ in the central parts of the LMC and around $0.1\times 10^{-7}$ when averaged over all OGLE-III fields.

The lenses should be dark in case of MACHO lensing, whereas in self-lensing the lens can be any star from the LMC.
Such a luminous lens can be indirectly ``detected'' when microlensing model indicates some amount of light blended in with the source.
For crowded fields such as those in the Magellanic Clouds observed with ground-based seeing, blending is a relatively common phenomenon.
In Section \ref{sec:blending} we tackled this issue with scrutiny and concluded that a single OGLE object is composed of about 1.2 real stars on average over the entire OGLE--III LMC coverage.
For the fields containing two our event candidates and for their respective magnitude range (Fig.\ref{fig:CF}) these factors are even smaller than average with 1.09 and 1.16 for LMC122.1 and LMC163.6, respectively. 
It means that there is less than about 15 per cent chance the blending detected in the microlensing model fit comes from random line-of-sight alignment of stars within a single seeing disk.

There is yet another tool we have at hand to confirm this last statement further.
Photometry processing using the DIA method provides also a precise position of the lensed flux with respect to the centroid of light measured on the template. The accuracy of such displacement measurement strongly depends on the magnitude at the peak and in the case of our events is at a level of 100 mas
(lensed images of the source are usually separated by less than 1 mas hence only centroid shift can be measured with such accuracy).
Because stars randomly blended together are usually displaced slightly from each other within the seeing disk, astrometric detection of a shift of the light barycentre provides an additional proof for ``natural'' random blending.
In the case of the two events presented here there was clearly no shift visible in astrometry of the residuals on the DIA subtracted images.
Unless we are very ``lucky'' and the blended star happen to be sitting exactly on the source of the event, this can be attributed to the fact that all the blending light comes from the lensing object. 
Hence, the lenses must be luminous. 
Their locations on the CMD, calculated from the blended microlensing model, indicate they belong to the LMC star locus. 
This clearly hints towards the self-lensing nature of the detected events.


The two events found manually are also difficult to attribute to MACHO lensing.
OGLE-LMC-05 is likely to be caused by a very red lens, probably located in the Galaxy disk.
OGLE-LMC-06 is generally very puzzling, with its light curve resembling a binary lens/source event and its very red colour, suggesting rather a variable star than a microlensing event.

It means we have no strong candidates for microlensing events caused by lenses from the halo.
We can, however, put an upper limit on the MACHO presence in the halo, similarly as done in \cite{TisserandEROSLMC}, who detected no events in their bright star sample. 
In our case, however, we have two events we associated with origin other than MACHO lensing.
This imply we can not apply straightforward zero-detection Poisson statistics, but should follow the suggestion of \cite{Moniez2010}, also applied in Paper II and treat our SL candidate events as an expected background. 
This is an obvious estimate, as more detailed studies involving LMC modelling are necessary in order to obtain exact amount of expected self-lensing events over the entire OGLE--III LMC sky coverage. 
Such analysis is planned to be performed in a way similar to the OGLE--II LMC study in \cite{CalchiNovati2009}.

\begin{figure*}
\includegraphics[width=13.5cm]{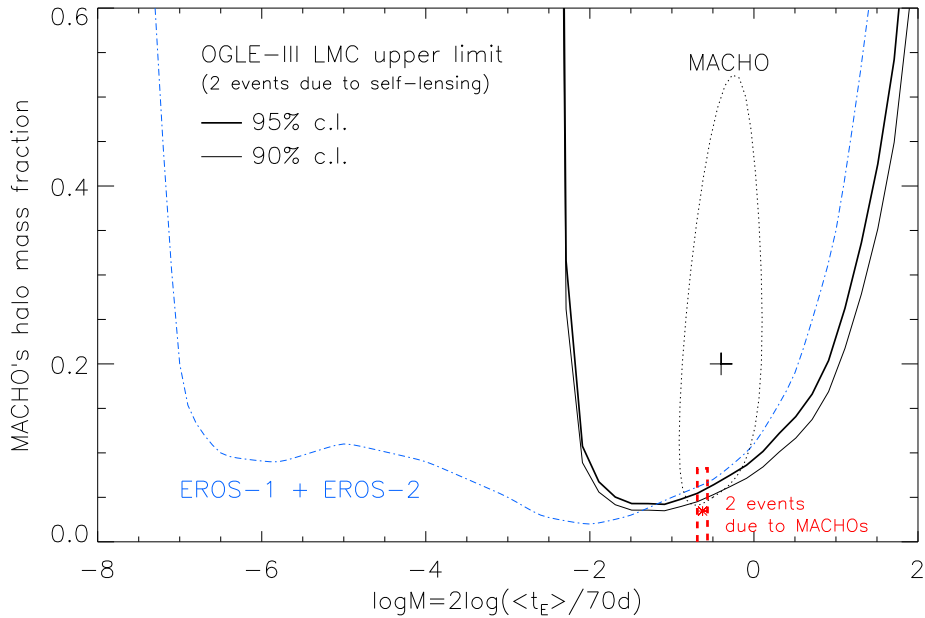}
\caption{
Fraction of the halo mass contained in the dark matter compact objects as derived from MACHO, EROS and OGLE-III data.
The red rectangle indicates the value for the case when both OGLE-III LMC events were caused by MACHO lenses.
Solid lines show upper limits for the case where there is no MACHO lensing event in the LMC OGLE--III data and both detected events are due to self-lensing.
MACHO curve (dotted line) denotes signal from the compact objects (at 95 per cent C.L.), whereas EROS curve (dot-dashed line) is an upper limit (at 95 per cent C.L.) on MACHO mass fraction in the halo.
}
\label{fig:taulimit}
\end{figure*}

Based on a mean detection efficiency over all fields we were able to estimate the number of expected events due to MACHOs considering model ``S'' of \cite{AlcockMACHOLMC}.
This number was calculated for a wide range of MACHO masses from $10^{-8}$ to $10^2~\msun$ and was translated to a fraction of halo mass using zero-statistics of \cite{FeldmanLowStatistics}.
It is shown in Fig. \ref{fig:taulimit}.
For masses around $M=0.4\msun$ we expected $N_{\rm exp}=69$ events for a halo full of MACHOs, it translates to an upper limit of  $f<7$ per cent at 95 per cent confidence and $f<6$ per cent at 90 per cent.
The limit reaches its minimum at around $M=0.1~\msun$ with $f<4$ per cent and is less rigid on masses higher than $0.4\msun$ reaching around 20 per cent at $M=10\msun$ and more at higher masses.

Our result is in agreement not only with previous LMC microlensing findings \citep{TisserandEROSLMC}, but also with studies of the microlensing effects of compact objects in distant galaxies observed in the lensed quasars (e.g. \citealt{Mediavilla2009}), which ruled out MACHOs in mass range between 0.1 and 10 $\msun$.
For the mass window $10-30~\msun$ there is still no reasonable constraint. 
There is actually some hint for heavy mass (around 10$\msun$) compact objects in the halo - \citet{Dong2007} studied the OGLE-2005-SMC-001 microlensing event and concluded it was caused by a binary black hole most likely located in the halo. 
However, it still remains a mystery why we don't see such events towards the LMC, therefore it is too early to conclude on dark matter compact objects existing in that mass window. 
OGLE--III SMC data, including that unique event, will be presented and studied in the forthcoming paper (Wyrzykowski et al. in prep.).

\section{Conclusions}

In this study we analysed almost 8 years of observations of the Large Magellanic Cloud by OGLE--III.
The data set with its volume, coverage and quality supersedes all previous determinations of the microlensing optical depth, including the one based on the OGLE--II data (Paper I).
We detected two sound candidates for microlensing events and further two possible candidates. 
Neither of them, however, is likely to be caused by dark matter compact lenses from the halo of our Galaxy.
The two best candidates can be explained as an expected signal from the self-lensing within the LMC. 
Of the remaining two, one is either a binary event or a some kind of chromatic outburst, whereas the other is a candidate for galactic disk lens.

Such null detection for MACHO lensing led to estimating the upper limit on their contribution to the mass of the Halo of the Galaxy.
The upper limit set at a level of 6-7 per cent at $M=0.4\msun$ leaves very little room for dark matter compact objects.
Still, at the moment we can not exclude more heavy dark matter lenses, like black holes. 
Our survey puts a 20 per cent halo mass fraction limit on compact objects with masses of $M=10\msun$ and actually no limit on higher masses.
This heavy mass end window should be now explored with more attention. 

As a side product of our analysis we also discovered that event MACHO-LMC-7, reported by the MACHO group and used in their final optical depth determination, exhibited couple of additional brightening episodes in the OGLE-III data, a feature which excludes it as a genuine microlensing event.

With the OGLE project continuing now in its fourth phase we hope the sensitivity to extremely long events will improve significantly within next years. 
It should result in the increase in the statistics of potential black-hole lenses or allow us to rule out heavy dark matter compact objects as well and close that topic definitively.


\section*{acknowledgements}
We would like to thank for their help at various stages of this work to Drs Nicholas Rattenbury, Vasily Belokurov and Patrick Tisserand.
We also thank the anonymous referee for their invaluable comments and remarks.
This work was partially supported by EC FR7 grant PERG04-GA-2008-234784 to {\L}W.
JS acknowledges support through the Polish MNiSW grant no. N20300832/0709 and Space Exploration Research Fund of The Ohio State University.
The OGLE project acknowledges funding received from the European Research Council under the European Community's Seventh Framework Programme (FP7/2007-2013), ERC grant agreement no. 246678. 

\bibliographystyle{mn2e}

\label{lastpage}

\end{document}